\documentclass[%
 reprint,
superscriptaddress,
groupedaddress,
nofootinbib,
 amsmath,amssymb,
 aps,
floatfix,
]{revtex4-2}

\usepackage{graphicx}
\usepackage{dcolumn}
\usepackage{bm}
\usepackage{hyperref}
\usepackage{mathtools,mathrsfs}%

\usepackage{xspace}
\usepackage{xcolor}
\usepackage{booktabs}
\usepackage{makecell}
\usepackage{multirow}

\newcommand{\GeV}{\,\text{GeV}\xspace}

\newcommand{\invfb}{\,\ensuremath{\mathrm{fb^{-1}}}\xspace}
\newcommand{\PW}{\ensuremath{W}\xspace}

\newcommand{\Vcb}{\ensuremath{|V_{cb}|}\xspace}
\newcommand{\ttbar}{\ensuremath{t\overline{t}}\xspace}
\newcommand{\Wjets}{\ensuremath{W}+jets\xspace}
\newcommand{\tW}{\ensuremath{tW}\xspace}
\newcommand{\WW}{\ensuremath{WW}\xspace}
\newcommand{\Dbc}{\ensuremath{D_{bc}}\xspace}

\newcommand{\hpm}{\ensuremath{H^{\pm}}\xspace}

\newcommand{\xbc}{\ensuremath{X\rightarrow bc}\xspace}
\newcommand{\xcs}{\ensuremath{X\rightarrow cs}\xspace}
\newcommand{\xbq}{\ensuremath{X\rightarrow bq}\xspace}
\newcommand{\xbqq}{\ensuremath{X\rightarrow bqq}\xspace}
\newcommand{\pt}{\ensuremath{p_{\text{T}}}\xspace}
\newcommand{\kt}{\ensuremath{k_{\text{T}}}\xspace}
\newcommand{\msd}{\ensuremath{m_{\text{SD}}}\xspace}
\newcommand{\Wcand}{\ensuremath{\PW_{\text{cand}}}\xspace}
\newcommand{\ptmiss}{\ensuremath{\pt^\text{miss}}\xspace}
\newcommand{\topbqq}{\ensuremath{\text{top}(bqq')}\xspace}
\newcommand{\topbq}{\ensuremath{\text{top}(bq)}\xspace}
\newcommand{\topbc}{\ensuremath{\text{top}(bc)}\xspace}
\newcommand{\Wqq}{\ensuremath{W(qq')}\xspace}

\newcommand{\MGvATNLO}{\texttt{MG5\_aMC@NLO}\xspace}
\newcommand{\PYTHIA} {{\texttt{Pythia}}\xspace}

\newcommand{\DELPHES} {{\texttt{Delphes}}\xspace}

\newcommand{\jetclassii} {{\text{JetClass-II}}\xspace}

\newcommand{\larger} {\text{large-$R$}\xspace}
\newcommand{\Sophon} {\texttt{Sophon}\xspace}
\newcommand{\SophonAKFour} {\texttt{SophonAK4}\xspace}

\begin{document}

\preprint{APS/123-QED}

\title{\boldmath Novel $|V_{cb}|$ extraction method via boosted $bc$-tagging with \textit{in-situ} calibration}

\author{Yuzhe Zhao}
\email[Electronic mail: ]{yuzhe.zhao@cern.ch}
\affiliation{School of Physics and State Key Laboratory of Nuclear Physics and Technology, Peking University, 100871 Beijing, China}

\author{Congqiao Li}
\email[Electronic mail: ]{congqiao.li@cern.ch}
\thanks{corresponding author}
\affiliation{School of Physics and State Key Laboratory of Nuclear Physics and Technology, Peking University, 100871 Beijing, China}

\author{Antonios Agapitos}
\affiliation{School of Physics and State Key Laboratory of Nuclear Physics and Technology, Peking University, 100871 Beijing, China}

\author{Dawei Fu}
\affiliation{School of Physics and State Key Laboratory of Nuclear Physics and Technology, Peking University, 100871 Beijing, China}

\author{Leyun Gao}
\affiliation{School of Physics and State Key Laboratory of Nuclear Physics and Technology, Peking University, 100871 Beijing, China}

\author{Yajun Mao}
\affiliation{School of Physics and State Key Laboratory of Nuclear Physics and Technology, Peking University, 100871 Beijing, China}

\author{Qiang Li}
\affiliation{School of Physics and State Key Laboratory of Nuclear Physics and Technology, Peking University, 100871 Beijing, China}

\date{\today}

\begin{abstract}
We present a novel method for measuring $|V_{cb}|$ at the LHC using an advanced boosted-jet tagger to identify ``$bc$ signatures''. By associating boosted $W \rightarrow bc$ signals with $bc$-matched jets from top-quark decays, we enable an \textit{in-situ} calibration of the tagger. This approach significantly suppresses backgrounds while reducing uncertainties in flavor tagging efficiencies---key to improving measurement precision. Our study is enabled by the development of realistic, AI-based large- and small-radius taggers, \texttt{Sophon} and the newly introduced \texttt{SophonAK4}, validated to match ATLAS and CMS's state-of-the-art taggers. The method complements the conventional small radius jet approach and enables a $\sim$30\% improvement in $|V_{cb}|$ precision under HL-LHC projections. As a byproduct, it enhances $H^{\pm} \rightarrow bc$ search sensitivity by a factor of 2--5 over the recent ATLAS result based on Run~2 data. Our work offers a new perspective for the precision $|V_{cb}|$ measurement and highlights the potential of using advanced tagging models to probe unexplored boosted regimes at the LHC.
\end{abstract}

\maketitle


\textbf{\textit{Introduction.---}}
Precise measurements of \Vcb are essential for understanding the Cabibbo-Kobayashi-Maskawa (CKM) sector of the standard model (SM). Established measurements have only relied on $B$-meson decay and have achieved an uncertainty of approximately 2\%~\cite{HFLAV:2022esi,HeavyFlavorAveragingGroupHFLAV:2024ctg}; yet, a persistent tension exists between results from inclusive and exclusive $B$ decay methods (see a review in Ref.~\cite{ParticleDataGroup:2024cfk}). Complementary extraction of \Vcb at the weak scale via $W$ boson decays through $\Gamma(W\to bc) / \Gamma(W\to qq')$~\cite{Harrison:2018bqi} provides a new angle to probe \Vcb, and with different conditions of theoretical and experimental uncertainties. Its potential has been explicitly explored in the context of HL-LHC~\cite{Azzi:2019yne} and future Higgs factories~\cite{Liang:2024hox}.
However, the small branching ratio of $W\to bc$ decays requires high-luminosity data to achieve competitive precision. A key challenge across these studies is controlling the experimental uncertainties from flavor-tagging and mistagging efficiencies~\cite{Harrison:2018bqi,Azzi:2019yne,Liang:2024hox}.
Reducing these uncertainties will be valuable to enhancing the unique opportunity to probe \Vcb via $W\to bc$ decays.

\begin{figure}[b]
\centering
\includegraphics[width=.40\textwidth]{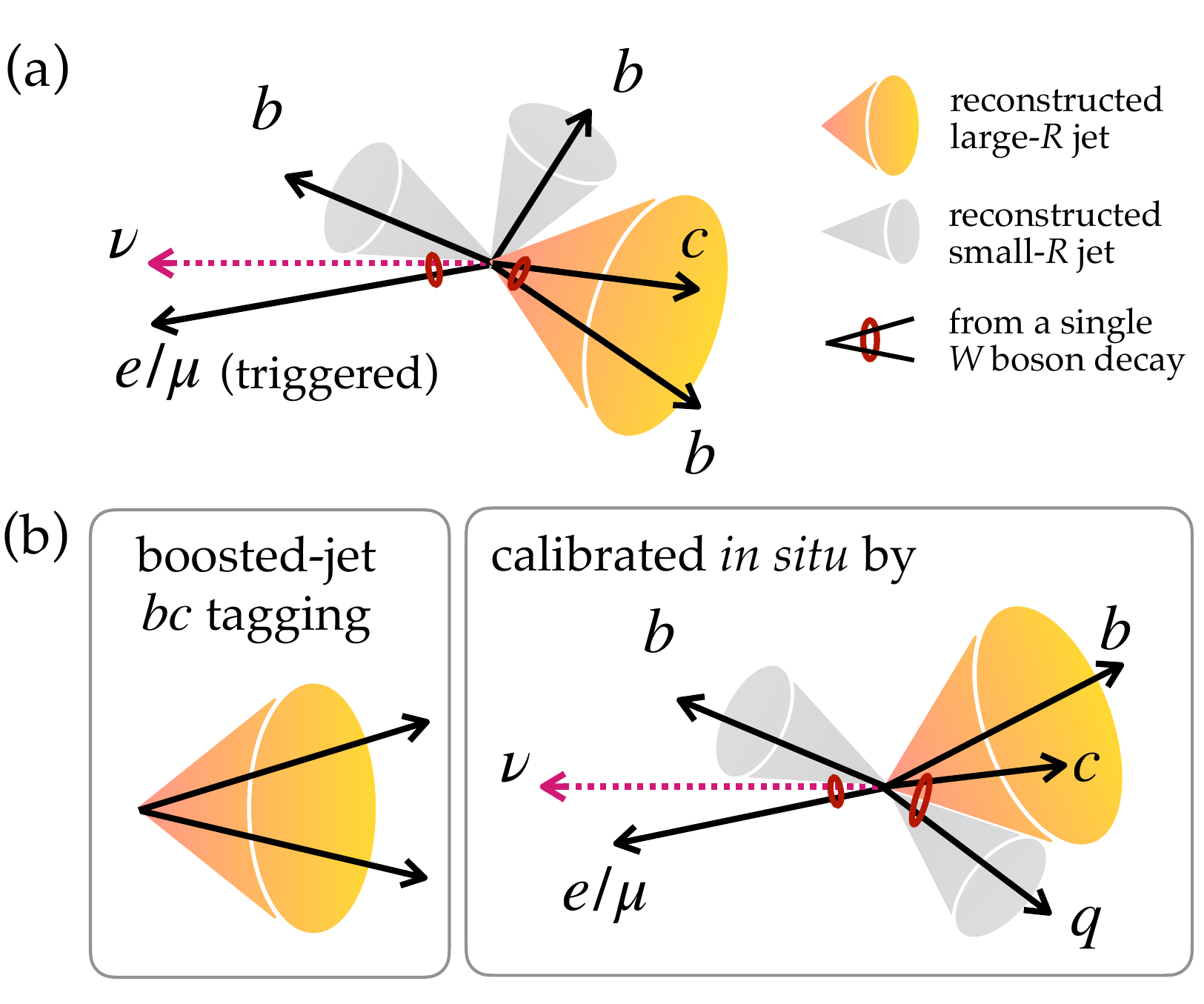}
\vspace{-5pt}
\caption{Illustration of (a) the boosted event topology of semi-leptonic \ttbar channel including a $W\to bc$ decay, and (b) techniques of boosted-jet $bc$ tagging and \textit{in-situ} calibration introduced in this work.
\label{fig:illustration}}
\end{figure}

We propose here a novel approach to extract \Vcb in the Lorentz-boosted regime at the LHC to mitigate this issue. This method offers two key advantages. First, boosted $bc$-tagging will enable significantly stronger suppression of background processes. This is similar to the recent advancements at the LHC, where increasingly sophisticated deep learning techniques for boosted $bb$- and $cc$-tagging have brought substantial improvements in measurements related to $H\rightarrow bb/cc$ decays~\cite{CMS:Hbb2016,CMS:HH4b,CMS:VHcc,CMS:ggHcc}. 
Second, a novel opportunity for employing $bc$-tagging is its ability to facilitate \textit{in-situ} calibration of the signal process, bypassing the need for calibration on a proxy phase space that typically introduces large uncertainties in tagging efficiencies as in $bb$ and $cc$ cases~\cite{CMS:JMETagger,ATLAS:bbTaggerCalibPaper}. This is because, in the semi-leptonic \ttbar phase space, hadronic top decays produce a distinct signature where a $b$ quark from $t$ decay and a $c$ quark from $W$ decay form one large-$R$ jet. With a stringent $bc$-tagging selection, the background is almost entirely dominated by ``$bc$-matched'' jets. Thus, the $bc$-tagger efficiency for the signal process can be corrected using a shared unconstrained scale factor derived from the above mentioned ``$bc$-matched'' background region.
This not only reduces uncertainty in boosted-jet $bc$-tagging efficiency but also mitigates the dependence of remaining event selections on $b/c$ flavor tagging, thereby enabling more precise extraction of \Vcb under high-luminosity conditions.
Figure~\ref{fig:illustration} highlights the key features of our proposed method.

We conduct in-depth experiments to demonstrate these advantages in this study. Our primary objective is to design synchronized strategies that enable a direct comparison between our novel boosted-channel approach with the established resolved-channel method, which only utilizes small-$R$ jet tagging (as in Refs.~\cite{Harrison:2018bqi,Choi:2021esu}).
To benchmark the advancements achieved in ATLAS and CMS's flavor tagging performance on large- and small-$R$ jets, we employ advanced deep learning models for tagging on \DELPHES simulation, the \texttt{Sophon} model~\cite{Li:2024htp} and the newly developed \texttt{SophonAK4}, which are validated to produce realistic performance as in actual experiments. Specifically, the \Sophon model is used to tag the resonance $bc$ signatures in the boosted regime.
Our synchronized strategies involve aligning the flavor tagger developments for large- and small-$R$ jets (via \texttt{Sophon} and \texttt{SophonAK4}), applying consistent multivariant techniques for event-level signal--background discrimination, and performing a unified counting analysis to extract \Vcb and evaluate the impact of flavor-tagging-related uncertainties.
In the HL-LHC projection, our results demonstrate that the uncertainty in \Vcb can be reduced by 30\% compared to the conventional approach, assuming consistent tagging performance benchmarks. The key factors driving this improvement are analyzed and discussed.
Furthermore, to complement this analysis, we benchmark the sensitivity of $t \rightarrow bH^{\pm} \rightarrow bbc$ search with the boosted-channel method. Our result suggests a 2--5-fold improvement in the upper limit on the product of branching fractions, $\mathscr{B}(t\to bH^{\pm}) \times \mathscr{B}(H^{\pm} \to bc)$, relative to the latest ATLAS results~\cite{ATLAS:2023bzb}, enabling a critical cross check of the reported $3\sigma$ excess.
Finally, we discuss the broader implications of this novel method and its impact on future LHC analyses.

\begin{figure}[htbp]
\centering
\includegraphics[width=.45\textwidth]{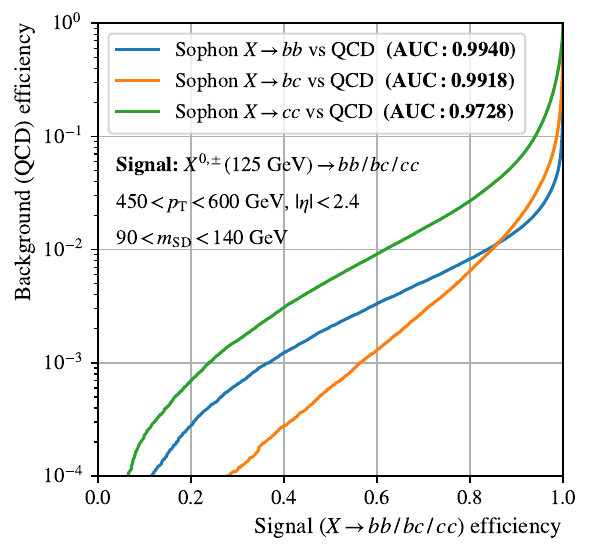}
\vspace{-10pt}
\caption{Performance of the \texttt{Sophon} model on $bb$, $bc$ and $cc$ tagging, shown as receiver operating characteristic (ROC) curves for signal jets versus the background QCD jets, with area under the curve (AUC) values annotated. Signal jets originate from resonant spin-0 $X^{0,\pm}\to bb/bc/cc$ decays with $m_X = 125\GeV$. The $bb$- and $cc$-tagging performance is consistent with CMS's state-of-the-art taggers~\cite{CMS-PAS-BTV-22-001,Li:2024jci} under the same phase-space selection.
\label{fig:roc-new}}
\end{figure}

\textbf{\textit{Flavor tagging models.---}}
Boosted $bc$-tagging has not been tested in actual LHC experiments but was covered in the recent \Sophon model proposed by some of the authors---a next-generation large-$R$ jet tagger designed for optimal performance across all possible boosted-jet final states~\cite{Li:2024htp}. It is essential to verify that the $bc$-tagging performance demonstrated by \Sophon aligns with what is achievable in actual experimental conditions.
Therefore, we compare resonance $bb$- and $cc$-tagging results against state-of-the-art CMS taggers~\cite{CMS-PAS-BTV-22-001,Li:2024jci} under consistency in signal and background definitions, jet selections, and discriminant definitions. Figure~\ref{fig:roc-new} benchmarks the performance of $bb$, $bc$, and $cc$ tagging, in terms of the receiver operating characteristic (ROC) curves and the area under the curve (AUC), with signal jets initiated from hypothetical spin-0 $X^{0,\pm}\to bb/bc/cc$ decay processes with $m_X=125$\GeV.
As expected, $bc$ tagging achieves an AUC between those of $bb$ and $cc$ tagging; at a tight working point, $bc$-tagging provides stronger QCD jet suppression compared to $bb$ and $cc$,  as QCD processes do not produce $bc$ from gluon splitting.
Additionally, $bb$- and $cc$-tagging results are found compatible with CMS's state-of-the-art taggers based on graph neural networks (\texttt{ParticleNet-MD}) and transformers (\texttt{GloParT}). See details for this comparison in Appendix~\ref{app:larger-tagger-perf}.

For conventional $b$ and $c$ taggers on small-$R$ jets, we develop the \SophonAKFour model in this work to synchronize the tagging techniques with \Sophon. The model is trained on anti-$k_\mathrm{T}$ small-$R$ jets ($R=0.4$)~\cite{Cacciari:2008gp} simulated with \MGvATNLO~v2.9.18~\cite{Alwall:2014hca} for event matrix-element generation, \PYTHIA~8.3~\cite{Sjostrand:2014zea} for hadronization, and \DELPHES~3.5~\cite{deFavereau:2013fsa} for fast detector simulation and object reconstruction, following the \jetclassii dataset configurations~\cite{Li:2024htp}. \SophonAKFour achieves $b$- and $c$-tagging performance compatible to ATLAS and CMS. Its $b$-tagging performance is slightly inferior to the widely-adopted \texttt{DL1r} tagger in ATLAS~\cite{ATLAS:FlvTag} and \texttt{DeepJet} in CMS~\cite{Bols:2020bkb}, and $c$-tagging performance falls between \texttt{DeepJet} and \texttt{ParticleNet} in CMS~\cite{Qu:2019gqs,CMS-DP-2024-066}. For choosing $b$- and $c$-tagging working points, we follow the recent ATLAS strategy for \texttt{DL1r}~\cite{ATLAS:VH}, defining five exclusive regions---\texttt{B1}, \texttt{B2} for $b$-tagged, \texttt{C1}, \texttt{C2} for $c$-tagged, and \texttt{N} for non-tagged regions---to facilitate simultaneous $b$- and $c$-jet identification and enable reasonable estimation of flavor-tagging uncertainties introduced below. Details on \SophonAKFour training, validation, and tagging region definitions are provided in Appendix~\ref{app:smallr-tagger-perf}.

\textbf{\textit{Experimental setup.---}}
Our experiment is conducted on simulated datasets for the LHC $pp$ collision at $\sqrt{s}=13$~TeV. We study the phase space triggered by a single isolated electron or muon. Thus, events are mainly dominated by the top quark-antiquark pair production (\ttbar) process, with contributions from other SM processes.
The simulated processes include \Wjets with leptonic $W$ decay, and \ttbar, single top in association with a $W$ boson (\tW) and diboson \WW production in their semi-leptonic decay channel. Events are generated with \MGvATNLO~v2.9.18 at the leading order (LO), with additional parton emissions matched to parton showers~\cite{Alwall:2007fs}, and decays of $t$ and $W$ implemented at the matrix-element level. The simulation employs SM values for $m_t$, $m_W$, and CKM matrix elements. The inclusive cross section of each process are scaled to their higher-order calculations~\cite{Czakon:2011xx,Melnikov:2006kv,Kidonakis:2012rm,Gehrmann:2014fva}.
Parton showers are modeled with \PYTHIA~8.3~\cite{Sjostrand:2014zea} with \texttt{NNPDF}~3.1 next-to-next-to-LO (NNLO) parton distribution function (PDF) set~\cite{Ball:2017nwa}.

The detector simulation is performed using \DELPHES~3.5 with the same configuration card as \jetclassii. It is based on the default CMS simulation card with modifications to (1) account for track smearing according to CMS tracker resolution, to enable realistic flavor tagging performance via \Sophon and \SophonAKFour on \DELPHES simulation, and (2) include pileup (PU) effects with an average of 50 PU vertices, and apply PU mitigation through the PU per particle identification (PUPPI) algorithm~\cite{Bertolini:2014bba}. The energy-flow (E-flow) objects modified by PUPPI are clustered into small- and large-$R$ anti-$k_{\mathrm{T}}$ jets with $R=0.4$ and $R=0.8$, using minimum \pt thresholds of 25 and 200\GeV, respectively. These procedures ensure consistency with algorithms used in CMS to mitigate PU and reconstruct jets.

The single-lepton trigger criteria are imposed, requiring a reconstructed electron with transverse momentum $\pt > 24$~GeV or a muon with $\pt > 32$~GeV, along with pseudorapidity and isolation constraints.
After trigger selection, the estimated event yields at an integrated luminosity of 140\invfb are $2.5\times 10^9$, $1.8\times 10^7$, $2.2\times 10^6$, and $3.2\times 10^6$ for \Wjets with leptonic $W$ decay, \ttbar, \tW, and \WW processes, respectively. The signal consists of the latter three processes with $W\to bc$ decays, yielding $1.8\times 10^4$ events in total under the same luminosity.

\textbf{\textit{Analysis strategies.---}}
We begin by describing the analysis strategy for the boosted regime.
In this context, events must contain at least one large-$R$ jet that is isolated from the trigger lepton, satisfying $\Delta R(\text{jet},\,\text{lepton}) > 0.8$. The jet with the highest \pt is designated as the $W$ candidate jet (\Wcand) and is required to have a soft-drop mass~\cite{Dasgupta:2013ihk,Larkoski:2014wba} within 60--110\GeV. Due to the topology of boosted events, only about 6\% of signal events survive this preselection. 
As illustrated by Fig.~\ref{fig:illustration}, the \Wcand jet can originate from the hadronic decay of a single $W$ or a top quark, containing all or part of their final-state products, or from QCD radiations. To differentiate these origins, $t$/$W$-initiated jets are categorized into five classes based on whether each daughter quark of $t$/$W$ is matched to the \Wcand jet within $\Delta R(\text{jet}, \text{quark}) < 0.8$. An additional category is defined for jets originating from QCD radiation. The six categories and their corresponding proportions are: \topbqq-matched (0.7\%), \topbc-matched (3.1\%), \topbq-matched (where $q$ is not a $c$) (12.3\%), \Wqq-matched (23.3\%), non-matched (14.7\%), and QCD-originated (45.9\%).

\begin{figure*}[!ht]
\centering
\includegraphics[height=.32\textwidth]{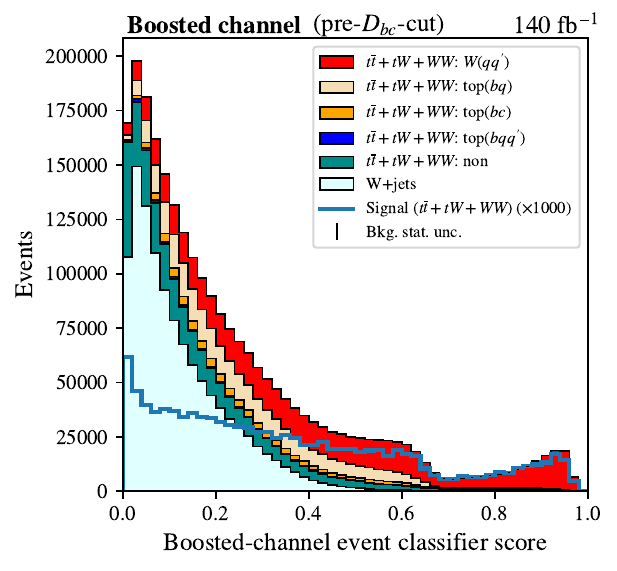}
\hspace{-15pt}
\includegraphics[height=.32\textwidth]{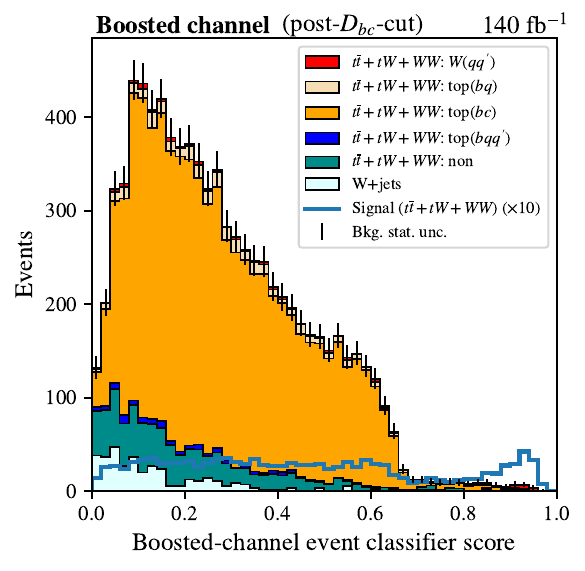}
\hspace{-15pt}
\includegraphics[height=.32\textwidth]{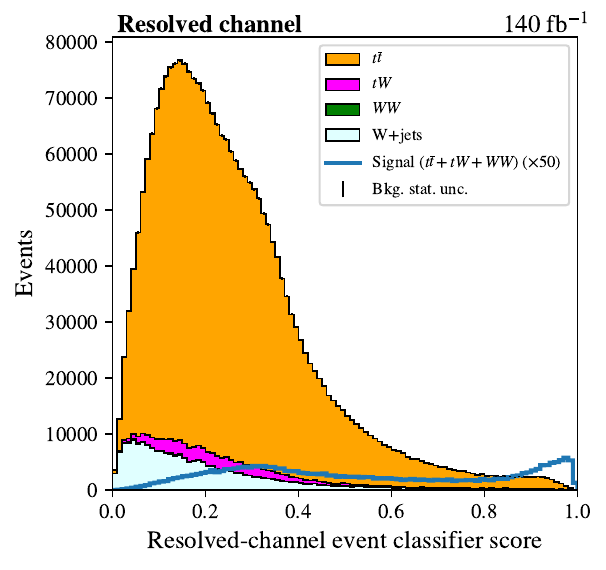}
\caption{\textbf{Left and middle}: Distributions of the classifier score in the boosted channel before and after the \Sophon discriminant \Dbc selection. \textbf{Right}: Distribution of the classifier score in the resolved channel. Note that in the boosted channel, the event-level classifiers aim to distinguish the \Wqq-matched jets from the ``top($bc$)+top($bq$)''-, \topbqq-, non-matched cases and \Wjets, while in the resolved channel, it is trained against the signal process against \Wjets, \ttbar, \tW, and \WW, background components, as elaborated in the text.
\label{fig:dist-evtscore}}
\end{figure*}

The boosted-regime method follows a two-stage strategy. First, a stringent selection is applied to the \Sophon discriminant to enhance the $bc$ content purity. Following the principle from Ref.~\cite{Li:2024htp}, the discriminant is defined as
\begin{equation}\label{eq:dbc}
    D_{bc} = \frac{g_{\xbc}}{g_{\xbc} + g_{\xbq} + g_{\xcs} + g_{\xbqq} + g_{\text{QCD}}}
\end{equation}
from \Sophon's 188 output scores $\Vec{g}$.
This formulation ensures that all background contributions are effectively suppressed.
In the second stage, a multivariate classifier is used to optimally distinguish \Wcand jets associated with \Wqq-matched signatures from other cases, without utilizing the \Sophon tagging information.
For this, we adopt an efficient deep learning approach: training a particle-transformer network~\cite{Qu:2022mxj} to classify events into the following categories: \Wqq-matched, ``top($bc$)+top($bq$)''-matched, \topbqq-matched, non-matched cases, and \Wjets background, using event-object-level features.
The input objects with their features include:
\begin{itemize}
\item The triggered lepton, with its 4-vector;
\item Missing transverse momentum \ptmiss, with its constructed 4-vector assuming $\eta$, $m=0$;
\item The \Wcand large-$R$ jet, with its 4-vector;
\item Up to five small-$R$ jets exclusive to the triggered lepton and \Wcand, with their 4-vectors and flavor-tagging labels, indicating the \SophonAKFour-tagged region (\texttt{B1}, \texttt{B2}, \texttt{C1}, \texttt{C2} and \texttt{N}) they belong to.
\end{itemize}
This deep learning algorithm efficiently combines kinematics properties, such as the invariant mass and $\Delta R$ of all possible object pairs, achieving superior classification performance~\cite{Qu:2022mxj}.
Figure~\ref{fig:dist-evtscore} (left and middle) shows the classifier output distribution for background events, categorized by their truth-matching criteria, and the signal process, both before and after the \Sophon \Dbc selection.
Notice that after a stringent \Sophon \Dbc selection, the background is predominantly composed of the \topbc-matched component.
We note that the spirit in this two-stage method is to use event-level information fully independent of \Wcand content to construct the classifier to decorrelate the classifier output with \Dbc. Thus, the \Dbc tagging efficiency can be calibrated by pairing the $bc$-matched background with the signal and correcting both yields using a shared, unconstrained scale factor, as will be detailed in the \Vcb extraction.

We then introduce the conventional resolved-regime strategy, similar to that initiated in Ref.~\cite{Harrison:2018bqi} while maintaining consistency with the boosted channel. Here, we require each event to have exactly one lepton and at least four small-$R$ jets exclusive to the triggered lepton, with at least three jets tagged as $b/c$ (i.e., labeled as \texttt{B1}, \texttt{B2}, \texttt{C1}, or \texttt{C2}). This selection retains about 28\% of triggered signal events. To achieve comparable event classification performance to the boosted regime, we employ a particle-transformer-based classifier to distinguish signal versus background events. The input objects (and features) for this classifier include the trigger lepton (its 4-vector), \ptmiss (its constructed 4-vector), and up to six small-$R$ jets exclusive to the triggered lepton (their 4-vectors and five \SophonAKFour tagging labels).
The classifier score distribution is shown in Fig.~\ref{fig:dist-evtscore} (right), showing the discrimination ability of signal events versus all background processes.
Further details on the classifier development for both the boosted and resolved channels can be found in Appendix~\ref{app:analysis}.

\textbf{\textit{\boldmath $|V_{cb}|$ extraction and uncertainty estimates.---}}
For both boosted and resolved channels, we extract \Vcb by performing a counting analysis on events that pass an optimized classifier score threshold.
We define the following event counts: $N_{\rm s}$ as the predicted signal events count, $N_{\rm b0}$ as the predicted count of backgrounds that do not contain a hadronically decayed \PW boson, and $N_{\rm b1}$ as the predicted count of backgrounds involving a hadronically decayed \PW boson from non-$bc$ sources.
The signal strength $\mu$ $(=N_{\rm s} / N_{\rm s}^{\rm SM})$ is related to \Vcb by
\begin{equation}\label{eq:mu}
 \mu = \left(\frac{|V_{cb}|^{\rm obs}}{|V_{cb}|^{\rm SM}}\right)^2,
\end{equation}
at leading order in $\alpha$ and $\alpha_s$. Notice that the inclusive hadronic \PW decay width is proportional to $\sum_{i=u,c;\,j=d,s,b} |V_{ij} |^2$, which is constant due to CKM unitarity. Thus, the sum $N_{\rm s} + N_{\rm b1}$ is constrained to be constant, leading to the relation
\begin{align}
 N_{\rm b1} & = \frac{1 - \mu r}{1 - r}N_{\rm b1}^{\rm SM},\\
 \text{where}\quad r & \equiv \frac{\Gamma(W\to bc)}{\Gamma(W\to qq')} = \frac{1}{2}(|V_{cb}|^{\rm SM})^2,
\end{align}
while $N_{\rm b0}$ remains unchanged. The likelihood function is constructed as a Poisson distribution with the mean value of $N_{\rm s}(\mu) + N_{\rm b1}(\mu) + N_{\rm b0}$, and is modified as follows to incorporate flavor-tagging-related uncertainties.

\begin{figure}[htbp]
\centering
\includegraphics[width=.48\textwidth]{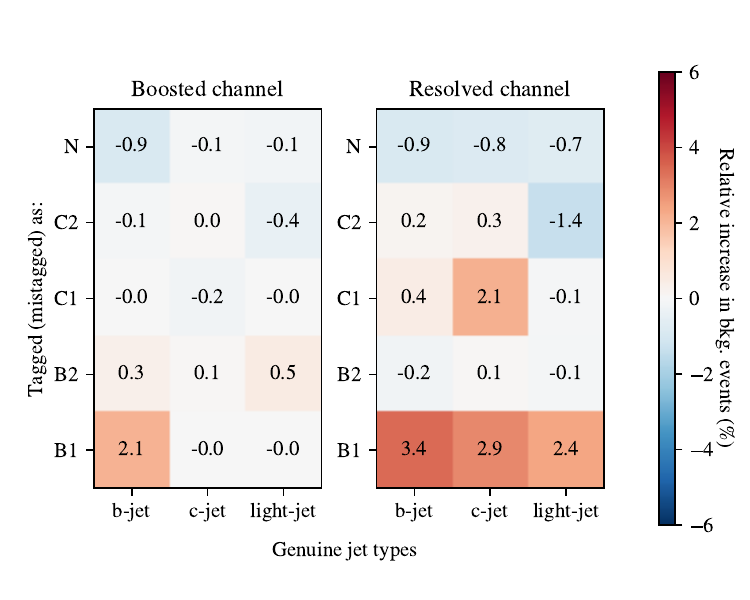}
\vspace{-10pt}
\caption{Comparison of the boosted and resolved channels showing the percentage increase in background events when each of the 15 flavor tagging and mistagging efficiencies is varied to $+1\sigma$. These 15 factors correspond to the efficiencies for genuine $b$, $c$, and light jets being tagged or mistagged into five categories: \texttt{B1}, \texttt{B2}, \texttt{C1}, \texttt{C2}, and \texttt{N}, as defined by the \SophonAKFour scores.
\label{fig:flavor_variation}}
\end{figure}

In both channels, we consider the effect of varying efficiencies for genuine $b$, $c$, and light jets tagged or mistagged into five regions (\texttt{B1}, \texttt{B2}, \texttt{C1}, \texttt{C2} and \texttt{N}), resulting in 15 independent factors affecting $N_{\rm s,b0,b1}$. In ATLAS and CMS, these efficiencies and their uncertainties are measured in dedicated $b$-, $c$- and light-flavor-enriched regions~\cite{ATLAS:btagSF,ATLAS:ctagSF,ATLAS:lightSF,CMS-DP-2023-005,CMS-DP-2023-006}. We reference the recent ATLAS measurement using the 140\invfb Run~2 data~\cite{ATLAS:VH} to acquire these uncertainties: the $b$- ($c$-)tagging efficiency uncertainty ranges from 0.01--0.08 (0.04--0.10), and the mistagging rate uncertainty for light jets as $b$ ($c$)-tagged jets ranges from 0.20--0.25 (0.12--0.13). Detailed values are provided in Appendix~\ref{app:smallr-tagger-perf}.
For each of the 15 sources, we vary the corresponding efficiency up to its $+1\sigma$ value, normalize the inclusive event yield to remain constant, then evaluate the impact on $N_{\rm s,b0,b1}$. This includes 15 nuisance parameters $\Vec{\nu}$ that modify the likelihood by each applying three multiplicative factors to scale $N_{\rm s,b0,b1}$, following log-normal distributions. We show in Fig.~\ref{fig:flavor_variation} the relative increase in the background count due to $+1\sigma$ variations in the 15 sources. The boosted and resolved channels are compared. We notice that in the boosted channel, $b$-tagging efficiency is a predominant factor, whereas in the resolved channel, multiple factors contribute more significantly to the overall uncertainty. These include both $b/c$-tagging efficiencies and $c$/light $\to$ $b$ jet mistagging rates. This is because the resolved-channel classifier relies more heavily on identifying multiple $b$ and $c$ jets; in the boosted channel, the $W\to bc$ tagging is integrated within \larger jet techniques. In addition, the classifier threshold is tighter in the resolved channel to optimize sensitivity. Figure~\ref{fig:flavor_variation} illustrates that the boosted channel is less affected by the small-$R$ jet flavor tagging uncertainties.

In the boosted channel, the large-$R$ jet \Dbc tagging efficiency is calibrated \textit{in situ} by introducing an unconstrained multiplicative factor $\lambda$, which is applied to both the signal and the \topbc-matched background in the post-\Dbc-cut region (i.e., the region without applying a cut on the classifier score).
This factor introduces an additional Poisson term in the likelihood to constrain the event yield in this region. Thus, $\mu$, $\Vec{\nu}$, and $\lambda$ will be simultaneously extracted in the fit.

\begin{figure}[htbp]
\centering
\includegraphics[width=.45\textwidth]{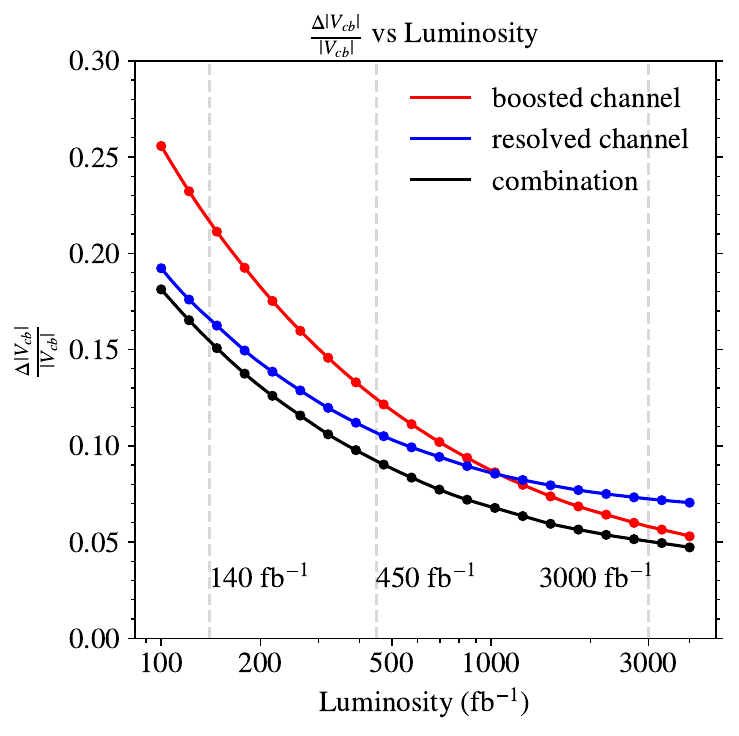}
\caption{The total fractional uncertainty of \Vcb obtained from the boosted and resolved channels, as well as their combination, presented under different luminosity conditions. Notably, in the HL-LHC scenario (3000\invfb), the precision achieved by the boosted channel surpasses that of the resolved channel.
\label{fig:results}}
\end{figure}

\textbf{\textit{Results and discussion.---}}
To summarize the \Vcb measurement results, Fig.~\ref{fig:results} presents the total uncertainty in \Vcb for the resolved and boosted channels under different integrated luminosity scenarios, obtained from the best-fit value and uncertainty of $\mu$. In the contexts of Run 2 (140\invfb), Run 2 and 3 combined (450\invfb), and the HL-LHC (3000\invfb), we also provide the uncertainties contributed by $b/c$ tagging and \Dbc tagging, and statistical errors, as listed in Table~\ref{tab:unc_breakdown}. These uncertainties are obtained by individually freezing $\Vec{\nu}$ and $\lambda$ in the fit. 
Here, we assume that the uncertainty in flavor tagging efficiency does not decrease with increasing luminosity, as in the actual experiments, the $b/c$ tagger scale factor measurements are often performed separately for different data-taking conditions~\cite{CMS-DP-2023-005,CMS-DP-2023-006,ATLAS:btagSF,ATLAS:ctagSF,ATLAS:lightSF}.
Notably, under the 140\invfb condition, our benchmark predicts an uncertainty for the conventional resolved channel as $\Delta \Vcb/\Vcb = 0.065\,(\text{flavor tag.~syst.})\oplus 0.154\,(\text{stat.})$. This is consistent with the preliminary expected result from ATLAS~\cite{Roberts:2882545} at around $0.13\,(\text{syst.})\oplus 0.13\,(\text{stat.})$, thus providing good validation to our flavor tagging benchmark and analysis strategy.

\begin{table}
\centering
\caption{Comparison of the fraction uncertainties of \Vcb contributed by small-$R$ jet flavor tagging (i.e., $b/c$ tagging), large-$R$ jet tagging (i.e, \Dbc tagging) and statistical errors, for the boosted and resolved channels.}
\vspace{5pt}
\begin{tabular}{lcccc}
\toprule
Luminosity (channel)\;\; & \;\;$b/c$-tag.\;\; & \;\;\Dbc-tag.\;\; & \;\;stat.\;\; \\
\midrule
140\invfb (boosted)  & 0.036 & 0.100 & 0.191 \\
140\invfb (resolved)  & 0.065 & --- & 0.154 \\
\midrule
450\invfb (boosted)   & 0.036 & 0.056 & 0.106 \\
450\invfb (resolved)  & 0.065 & --- & 0.086 \\
\midrule
3000\invfb (boosted)   & 0.035 & 0.022 & 0.041 \\
3000\invfb (resolved)  & 0.065 & --- & 0.033 \\
\bottomrule
\end{tabular}
\label{tab:unc_breakdown}
\end{table}

From Fig.~\ref{fig:results} and Table~\ref{tab:unc_breakdown}, we observe that the boosted channel is predominantly limited by statistical uncertainty, while the contribution from flavor-tagging-related uncertainty is smaller than that of the resolved channel. As the luminosity increases, the boosted channel demonstrates a significant advantage in the overall uncertainty, surpassing the traditional resolved approach. We attribute this reduced flavor-tagging-related uncertainty to two key factors. First, the use of a novel calibration strategy for the \Dbc tagger avoids the complexity of finding intricate phase spaces for calibration as in the case of $bb$- or $cc$-taggers. Instead, it exploits local \textit{in-situ} constraints for calibration. Second, the decoupling of the event classifier from \Dbc tagging in the boosted strategy mitigates the classifier's sensitivity to the change in flavor tagging efficiencies.  Consequently, although the boosted regime targets only a small fraction of $W\to bc$ decays, its better control over uncertainties allows it to outperform the traditional resolved method at the HL-LHC condition.

Given that only around 15\% of the events in the resolved channel overlap with those in the boosted channel, a combined measurement can be achieved after excluding these overlapping events from the resolved channel. A simultaneous fit is thus performed, allowing the nuisance parameters to impact both channels. Under the HL-LHC (3000\invfb) scenario, this combined approach achieves a precision of $\Delta \Vcb/\Vcb = 0.051$, representing a 30\% reduction in uncertainty compared to the traditional resolved method. The performance of the combined scenario across different luminosities is also showcased in Fig.~\ref{fig:results}. 
Finally, a combination of measurements from both ATLAS and CMS experiments can further reduce the fraction uncertainty to 0.036, a level of precision sufficient to provide critical insights into resolving the \Vcb puzzle.

Although our study demonstrates the potential of the boosted regime in LHC \Vcb measurement, the precision of the $b/c$ tagging efficiency measurement remains a primary limiting factor. In an auxiliary investigation, we observed that under a more optimistic scenario, where the associated uncertainty is reduced by a factor of 2, the expected improvement in \Vcb precision at 3000\invfb diminishes from 30\% to 15\%.
Besides, we have only considered a limited set of systematic uncertainties. In actual experiments, additional systematic uncertainties arise from experimental sources (e.g., jet energy calibration) and theoretical modeling inaccuracies, which affect both the boosted and resolved regimes. Furthermore, precise control of other background processes is critical. One can rely on multiple control regions to accurately constrain the background contributions.

\begin{figure}[htbp]
\centering
\includegraphics[width=.45\textwidth]{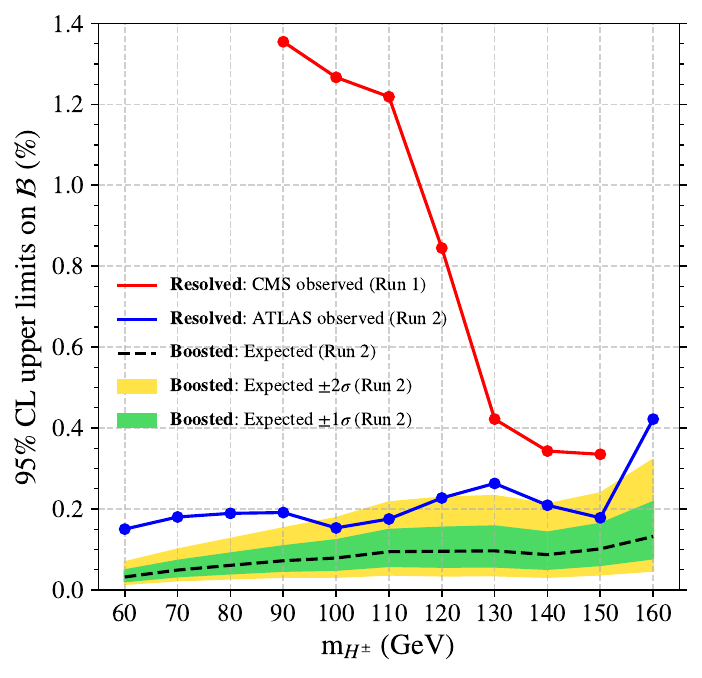}
\vspace{-10pt}
\caption{Expected 95\% CL upper limits on $\mathscr{B} = \mathscr{B}(t\to bH^{\pm}) \times \mathscr{B}(H^{\pm} \to bc)$ as a function of $m_{H^{\pm}}$. The green (yellow) band represents the $\pm 1\sigma$ ($\pm 2\sigma$) uncertainty range under the background-only hypothesis. Compared to existing constraints from CMS (red) and ATLAS (blue), our expected results improve upon the latest ATLAS Run~2 limits by a factor of 2--5.
\label{fig:hbc_upper_limit}}
\end{figure}

\textbf{\textit{Benchmarking $\hpm\to bc$ search.---}}
As a byproduct of this analysis, we benchmark a constraint on the $t \to bH^{\pm} \to bbc$ branching fraction using a boosted search strategy similar to that in the \emph{Analysis strategies} section. This process has been previously explored by CMS (with 19.7\invfb of Run~1 data)~\cite{CMS:2018dzl} and ATLAS (139\invfb of Run~2 data)~\cite{ATLAS:2023bzb}. Signal samples for $\ttbar \to bH^{\pm} b W^{\mp}$ are generated with a similar \MGvATNLO setup as $\ttbar$ backgrounds, with $m_{H^{\pm}}$ ranging from 60 to 160\GeV, a 1\GeV width, and decays entirely to $bc$, under a consistent model setup as in Ref.~\cite{ATLAS:2023bzb}. The leading boosted jet is identified as the ``$H^{\pm}$ candidate'', applying the same \Sophon \Dbc selection as defined in Eq.~(\ref{eq:dbc}). The event classifier is re-optimized by redefining the \Wqq-matched category as an $H^{\pm}(qq’)$-matched category, with training samples taken from a dedicated $\ttbar \to bH^{\pm} b W^{\mp}$ simulation with variable $H^{\pm}$ mass.
A binned-template likelihood fit is performed on the \msd distribution in four \pt regions (200--250, 250--300, 300--400, $>$400\GeV) to account for the kinematics of different $m_{H^{\pm}}$ values. Systematic uncertainties in $b/c$-flavor tagging and \Dbc tagging are incorporated via $\nu$ and $\lambda$ parameters, respectively. We extract the expected 95\% CL upper limit on $\mathscr{B} = \mathscr{B}(t\to bH^{\pm}) \times \mathscr{B}(H^{\pm} \to bc)$ under the background-only hypothesis using the $\mathrm{CL}_s$ method~\cite{Read:2002hq}, finding a 2--5-fold improvement over previous ATLAS results, presented in Fig.~\ref{fig:hbc_upper_limit}.
Notably, ATLAS has reported a local $3\sigma$ excess at $m_{H^{\pm}} = 130$ GeV. Given that our search explores a largely distinct boosted $H^{\pm}$ phase space and is capable of providing a more stringent constraint, it will offer a critical cross-check of this excess.

\textbf{\textit{Conclusion.---}}
In summary, we propose a novel approach for precise \Vcb extraction at the LHC, utilizing the boosted regime. By employing an advanced boosted-jet $bc$-tagger via the \Sophon model and an innovative \textit{in-situ} calibration technique, we factorize the complex requirements for event-level $b/c$ tagging and global calibration into two components: an integrated $bc$ tagging and calibration, and a residual component that is less sensitive to uncertainties in $b/c$ tagging and mistagging efficiencies. This effectively mitigates the impact of uncertainties related to flavor tagging, achieving a generally improved precision over the conventional method.
As a byproduct, this strategy also benchmarks the $H^{\pm} \to bc$ search in the $\ttbar$ phase space, achieving a large improvement over the latest ATLAS results.

From a broader sense, recent advancements in deep learning algorithms for particle physics have enhanced sensitivity through well-established boosted channels such as $bb$ and $cc$, revealing potential in exploring a broader range of boosted final states. This highlights the \Sophon philosophy: extending boosted-jet techniques to previously unexplored regions. Hence, this work not only demonstrates an application in a novel boosted phase space beyond established $bb$ and $cc$ contexts but also suggests a more important role for boosted-regime searches in future LHC explorations. As benchmark tagging models with consistent performance on \DELPHES datasets, \Sophon and \SophonAKFour will enable future investigations into diverse boosted phase spaces.
\vspace{10pt}

\textbf{\textit{Acknowledgement.---}}
This work is supported by National Natural Science Foundation of China (NSFC) under Grants No.~12325504, No.~12061141002, and No.~2075004.
YZ and CL appreciate helpful discussions with Sitian Qian and Huilin Qu.
This research is supported in part by the computational resource operated at the Institute of High Energy Physics (IHEP) of the Chinese Academy of Sciences.

\textbf{\textit{Data availability.---}}
The \Sophon model (for large-$R$ jets), \jetclassii dataset, and its corresponding \DELPHES configuration initiated in Ref.~\cite{Li:2024htp} are publicly available at the following https URLs\footnote{\url{https://huggingface.co/jet-universe/sophon}}\footnote{\url{https://huggingface.co/datasets/jet-universe/jetclass2}}.
The \SophonAKFour model will be made publicly available at this URL\footnote{\url{https://huggingface.co/jet-universe/sophon-ak4}}.

\clearpage
\onecolumngrid
\appendix
\renewcommand{\thefigure}{\Alph{section}\arabic{figure}}
\renewcommand{\thetable}{\Alph{section}\arabic{table}}
\setcounter{figure}{0}
\setcounter{table}{0}

\section{\boldmath Supplementary details on \texttt{Sophon}'s \texorpdfstring{$bc$}{bc}-tagging performance}\label{app:larger-tagger-perf}

During the training of the \Sophon model, the $bb$, $bc$, and $cc$ jet classes are generated through the decay of a spin-0 resonance $X$ into $bb$, $bc$, and $cc$ final states. These jets exhibit a broad distribution in \msd and \pt: $200 < \pt < 2500\GeV$ and $20 < \msd < 500\GeV$. For these jets, a $\Delta R$ matching condition between the jet and the two daughter quarks is imposed. During training, sampling-based reweighting is performed on the three jet classes and the QCD jets (as background), along with other categories. This involves a two-dimensional reweighting on $(\msd,\,\pt)$ to ensure consistent distributions, hence reducing the dependence of the tagger response on jet mass and \pt. 
Notice that the technical details for the $bb$ and $cc$ jets are consistent with the \texttt{GloParT} training method within CMS~\cite{Li:2024jci}, and the definition of the $bc$ jet class is similar to that of $bb$ and $cc$, which differs only in quark flavor. Therefore, by verifying the consistency of $bb$ and $cc$ tagging with actual experimental results, we can reasonably conclude that the performance of the Sophon $bc$ tagger is comparable to that observed in actual experiments.

The performance evaluation of \Sophon (Fig.~\ref{fig:roc-new}) is conducted on a $125 \GeV$ resonance $X^{0,\pm}$ decaying into $bb$, $bc$, and $cc$ as signal jets. The kinematic properties of the resonance $X$ follow the $\pt$ and $\eta$ distributions observed in gluon-gluon-fusion Higgs boson production. It is verified that the kinematic distributions of $bb$, $bc$, and $cc$ jets are consistent. The event selection requires the presence of a reconstructed large-$R$ jet, with the leading jet chosen as the candidate. The selected jet must satisfy $450 < \pt < 600$, $|\eta|<2.4$, and a soft-drop mass requirement of $90 < m_{SD} < 140$. These sample settings and kinematic selections are consistent with those used in the experiment. For signal jets, a jet-quark matching condition is also required. 
The QCD background is obtained from QCD multijet events, where the leading large-$R$ jet is required to satisfy the same kinematic requirements. The definition of the \Sophon discriminant follows the same convention as \texttt{ParticleNet-MD} and \texttt{GloParT}. It is defined as
\begin{equation}
\text{discr~($X\to$~sig vs QCD)} = \frac{g_{X\to\text{sig}}}{g_{X\to\text{sig}} + g_{\rm QCD}},\quad\text{where sig~$= bb$, $bc$, or $cc$}.
\end{equation}

Table~\ref{tab:bkgrej_sophon} presents the background rejection rate ($=1/\epsilon_B$) for $bb$ and $cc$ discriminants at signal efficiencies of $\epsilon_S = 60\%$ and $40\%$, compared with CMS $bb/cc$ taggers: \texttt{DeepDoubleX}, \texttt{ParticleNet-MD} (read from Ref.~\cite{CMS-PAS-BTV-22-001}, Figs.~5--6), and \texttt{GloParT} stage-2 (read from Ref.~\cite{Li:2024jci}, Figs.~11.12--13). The rejection rate of \texttt{GloParT} stage-2 is obtained by taking the ratio of the \texttt{GloParT}-to-\texttt{ParticleNet-MD} rejection rate from Ref.~\cite{Li:2024jci} and applying it to the \texttt{ParticleNet-MD} results from Ref.~\cite{CMS-PAS-BTV-22-001} due to a slight difference in their \pt selections. 
It is observed that for the $bb$ tagger, \texttt{Sophon}'s performance is between the CMS taggers \texttt{DeepDoubleX} and \texttt{ParticleNet-MD}, but closer to the latter. For the $cc$ tagger, \Sophon shows superior performance, nearly reaching the best results of \texttt{GloParT} in CMS. 
This difference in $b/c$ tagging performance can be attributed to discrepancies in the $b$ and $c$ jets simulation by the \DELPHES JetClass-II card compared to the actual detector response, as well as the absence of auxiliary variables, such as those related to tracks and secondary vertices, which can further enhance $b$ tagging performance.
Specifically, the result indicates that $b$ tagging performance may be slightly underestimated, as also reflected in the performance of \SophonAKFour in the following appendix section.

\begin{table}[!ht]
\centering
\caption{Comparison of the \Sophon model with the established CMS taggers, \texttt{DeepDoubleX}, \texttt{ParticleNet-MD}, and \texttt{GloParT} (stage-2), regarding $bb$ and $cc$ tagging performance. The QCD background rejection rate ($1/\epsilon_B$) is shown for $bb$ and $cc$ tagging at fixed signal efficiencies of $\epsilon_S = 60\%$ and $40\%$. The CMS results are extracted from the ROC curves in Refs.~\cite{CMS-PAS-BTV-22-001,Li:2024jci}, as detailed in the text.}
\vspace{5pt}
\label{tab:bkgrej_sophon}
\begin{tabular}{lcccc}
\toprule
  & \DELPHES simulation & \multicolumn{3}{c}{CMS simulation~\cite{CMS-PAS-BTV-22-001,Li:2024jci}} \\
\cmidrule(lr){2-2} \cmidrule(lr){3-5}
  & ~~\Sophon~~ & ~~\texttt{DeepDoubleX}~~ & ~~\texttt{ParticleNet-MD}~~ & ~~\texttt{GloParT} (stage-2) \\ 
\midrule
$X\to bb$ vs.~QCD, $\epsilon_{\rm s} = 60\%$ & 300 & 200 & 370 & 470 \\ 
$X\to bb$ vs.~QCD, $\epsilon_{\rm s} = 40\%$ & 810 & 580 & 970 & 1380 \\ 
\midrule
$X\to cc$ vs.~QCD, $\epsilon_{\rm s} = 60\%$ & 110 & 31 & 76 & 110 \\ 
$X\to cc$ vs.~QCD, $\epsilon_{\rm s} = 40\%$ & 320 & 110 & 260 & 360 \\ 
\bottomrule
\end{tabular}
\end{table}

\setcounter{figure}{0}
\setcounter{table}{0}

\section{Supplementary details on \texttt{SophonAK4} development}\label{app:smallr-tagger-perf}

\subsection{The training of \texttt{SophonAK4}}

To generate the training jet samples for \SophonAKFour, we employ the spin-0 resonance process $X$ decaying into multiple two-prong states. This is consistent with the production mechanism used in \Sophon training for two-prong large-$R$ jets.
Specifically, events are generated using the $p\,p\to H\,H$ process at LO with the \MGvATNLO~v2.9.18 generator, utilizing the \textsc{heft} model. To control the \pt and mass of the resonant jets, the minimum \pt of the $H$ boson at the hard-scattering level is sampled at 50 logarithmically spaced points between $(100,\,2500)$\GeV. The mass of the $H$ boson is uniformly sampled from $(15,\,500)$\GeV with an interval of 5\GeV.
The decay of the $H$ resonance and parton showering are simulated using \PYTHIA 8.3. The decay modes (with their corresponding branching ratios) are as follows: $bb~(\frac{1}{8})$, $cc~(\frac{1}{8})$, $ss~(\frac{1}{8})$, $dd~(\frac{1}{16})$, $uu~(\frac{1}{16})$, $gg~(\frac{1}{8})$, $ee~(\frac{1}{16})$, $\mu\mu~(\frac{1}{16})$, $\tau\tau~(\frac{1}{4})$.

The simulated events are processed using \DELPHES 3 with the \jetclassii configuration. As introduced in Ref.~\cite{Li:2024htp}, this simulation card is adapted from the CMS detector configuration but includes modifications to the impact parameter of charged particles to match the CMS tracker resolution, following the approach used in the JetClass simulation.
Additionally, PU effects with an average of 50 PU interactions are incorporated using the CMS detector configuration with PU. To mitigate PU contamination, the PU per-particle identification (PUPPI) algorithm is applied. This implementation is adapted from the CMS Phase-II detector configuration but includes parameter adjustments to match the Phase-I CMS detector conditions.
Specifically, PUPPI assigns a probability value between 0 and 1 to each E-flow object, indicating the likelihood that the object originates from the primary interaction. This value is then used to scale the object's four-vector.
The processed E-flow objects are clustered into small-$R$ jets using the anti-\kt algorithm with $R=0.4$.

Based on truth-matching information, each jet is assigned one of the single-prong or two-prong labels, resulting in a total of 23 jet categories:
\begin{enumerate}
\item The single-prong labels include $b$, $\overline{b}$, $c$, $\overline{c}$, $s$, $\overline{s}$, $d$, $\overline{d}$, $u$, $\overline{u}$, $g$, $e^-$, $e^+$, $\mu^-$, $\mu^+$, $\tau_{\rm h}^-$, and $\tau_{\rm h}^+$. These correspond to cases where the corresponding truth particle (either a parton or a lepton) is matched to the jet within $\Delta R (\text{jet, particle}) < 0.4$, and the other particle from the same resonance decay is not matched to the jet.
\item The two-prong labels include $b\overline{b}$, $c\overline{c}$, $s\overline{s}$, $d\overline{d}$, $u\overline{u}$, and $gg$, corresponding to cases where both particles from the same resonance decay are matched within the jet.  
\end{enumerate}
Jets are required to satisfy the kinematic criteria of $15 < \pt < 1000\GeV$ and $|\eta| < 5$. In total, 46~M jets are used for training the model.  

For each jet, only jet-consistent-level E-flow features are used as input variables. These features include kinematic variables, particle identification variables, and impact parameter features. This approach closely follows the JetClass dataset configuration and is consistent with the training of the \Sophon model, except that small-$R$ jets are used in this case.  

Similar to \Sophon, the \SophonAKFour model adopts the Particle Transformer architecture but with a reduced scale: the embedding dimension and the latent space size of the corresponding multilayer perceptrons (MLPs) are reduced by half compared to the original \Sophon model.
Specifically, \SophonAKFour consists of 8 particle attention blocks and 2 class attention blocks, with an embedding dimension of 64 and 8 attention heads. The initial particle features are embedded using a 3-layer MLP with (64,\,256,\,64) nodes, while the pairwise particle features are embedded using a 4-layer elementwise MLP with (32,\,32,\,32,\,8) nodes. The \texttt{GELU} activation function is used throughout the model. Overall, the \SophonAKFour model includes around 0.55~M parameters.

A sampling-based reweighting strategy is employed on the two-dimensional histogram of $(\pt,\,\eta)$ to ensure uniform classification performance across all \pt and $\eta$ regions. Specifically, training samples are selected into the training pool with predefined probabilities during the on-the-fly data loading process. These probabilities act as reweighting factors, adjusting the two-dimensional histograms bin by bin.
The reweighting is designed to produce normalized distributions for specific reweighting classes. These classes are formed by merging the 23 finely classified categories into 9 groups according to their parton or lepton flavors: $b$, $c$, $s$, $d$, $u$, $g$, $e$, $\mu$, and $\tau_{\rm h}$.

The model is trained using a batch size of 512 with an initial learning rate of $5\times 10^{-4}$. The full dataset is split into 80\% for training and 20\% for validation.
The training process spans 80 epochs, with each epoch processing 10~M samples. The optimizer and learning rate scheduler are identical to those used in the \Sophon model training.

\subsection{Performance of \SophonAKFour}

The performance evaluation of \SophonAKFour is conducted on \ttbar events from SM processes to facilitate direct comparison with the performance benchmarks from ATLAS and CMS.

We first compare the performance using ROC curves against various CMS taggers. To ensure consistency, we apply the same phase space selection criteria: jets are required to satisfy $\pt > 30\GeV$ and $|\eta|<2.5$. Jets from \ttbar events are assigned truth flavor labels as genuine $b$, $c$, and light jets using ghost association in \DELPHES. We define the following $b$-tagging discriminant to evaluate the performance of $b$ vs.~light jets and $b$ vs.~$c$ jets,
\begin{equation}
\text{discr~(\SophonAKFour $b$ tagging)} = g_{b} + g_{\overline{b}} + g_{b\overline{b}}.
\end{equation}
where $\Vec{g}$ represents the raw output scores of the model.  
For $c$-tagging, we define the following discriminants,
\begin{align}
\text{discr~(\SophonAKFour $c$ tagging)} & = g_{c} + g_{\overline{c}} + g_{c\overline{c}},\\
\text{discr~(\SophonAKFour $c$ vs.~$b$ tagging)} & = \frac{g_{c} + g_{\overline{c}} + g_{c\overline{c}}}{g_{c} + g_{\overline{c}} + g_{c\overline{c}} + g_{b} + g_{\overline{b}} + g_{b\overline{b}}}.
\end{align}
These are used to evaluate the performance of $c$ vs.~light jets and $c$ vs.~$b$ jets, respectively, following the conventions typically used by CMS~\cite{CMS-DP-2024-066}.  

\begin{figure}[htbp]
\centering
\includegraphics[width=.45\textwidth]{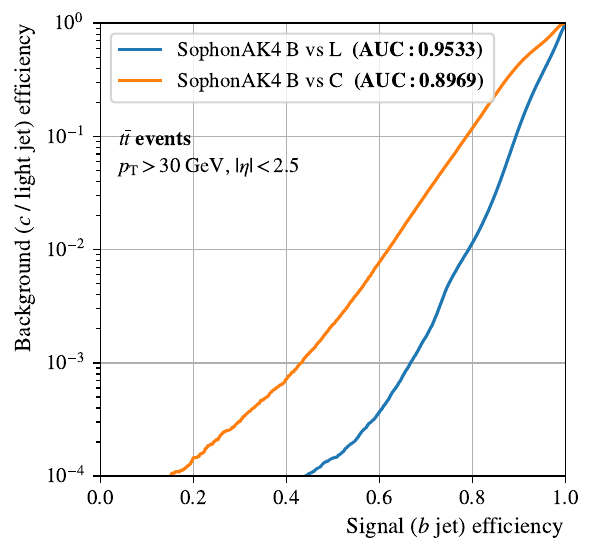}
\includegraphics[width=.45\textwidth]{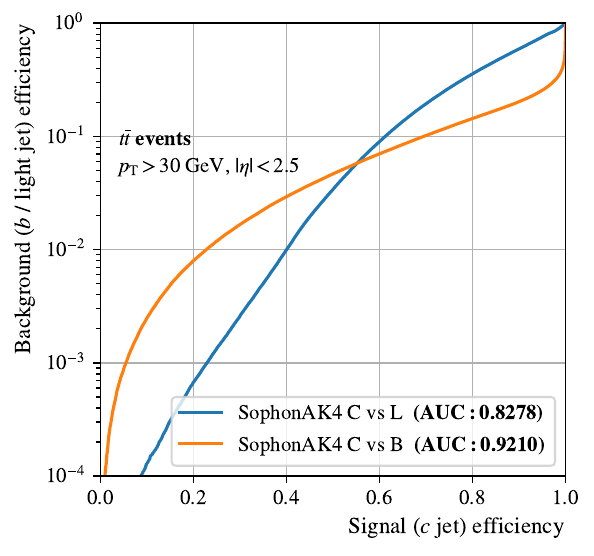}
\caption{Performance of the \texttt{SophonAK4} model for $b$ versus light/$c$ jet tagging (left) and $c$ versus light/$b$ jet tagging, shown as receiver operating characteristic (ROC) curves for signal jets versus the background QCD jets, with area under the curve (AUC) values annotated. The signal and background jets originate from \ttbar events and are labeled as genuine $b$, $c$, and light jets based on the ghost association criteria. These benchmarks are directly comparable to the performance of the CMS $b$ and $c$ flavor taggers under the same phase-space selection~\cite{CMS-DP-2024-066}, as detailed in the text.
\label{fig:roc_sophonak4}}
\end{figure}

Figure~\ref{fig:roc_sophonak4} presents the performance of $b$ vs.~light/$c$ jets and $c$ vs.~light/$b$ jets. Comparing these results with the CMS benchmarks (refer to Ref.~\cite{CMS-DP-2024-066}, Figs.~1 and 3 for the \ttbar process), we observe that the $b$ vs.~light jet performance is slightly inferior to the widely-adopted \texttt{DeepJet}, while the $b$ vs.~$c$ and $c$ vs.~light/$b$ jet performances lie between those of \texttt{DeepJet} and \texttt{ParticleNet}.  
This observation is consistent with the findings in Appendix~\ref{app:larger-tagger-perf} regarding the performance of \Sophon. Specifically, the \DELPHES simulation tends to yield weaker $b$ tagging capability but provides a more realistic $c$ tagging performance.  

\begin{figure}[htbp]
\centering
\includegraphics[width=.40\textwidth]{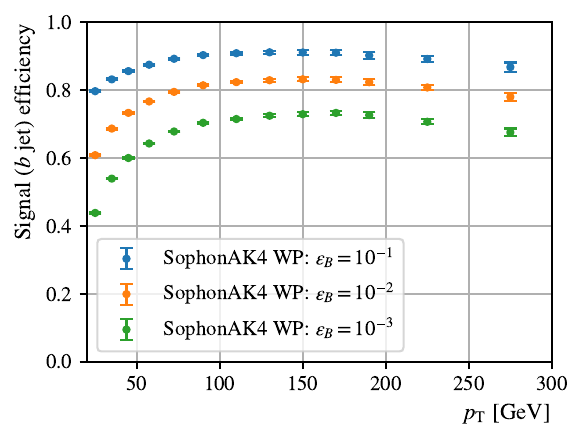}
\includegraphics[width=.40\textwidth]{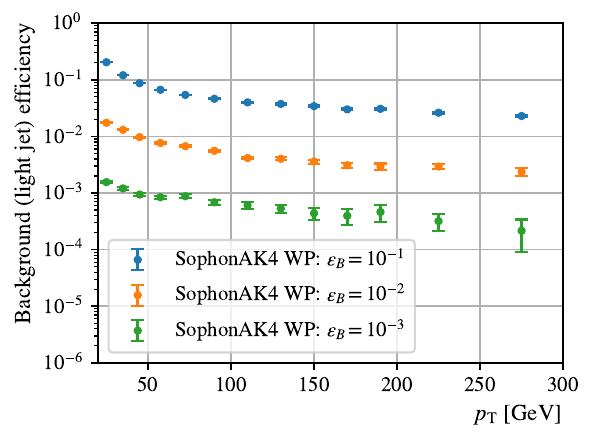}\\
\includegraphics[width=.40\textwidth]{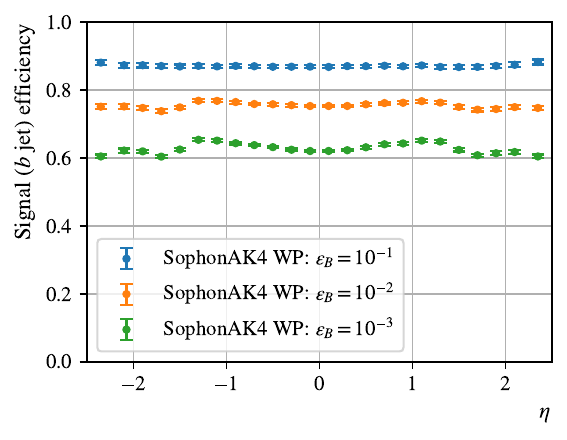}
\includegraphics[width=.40\textwidth]{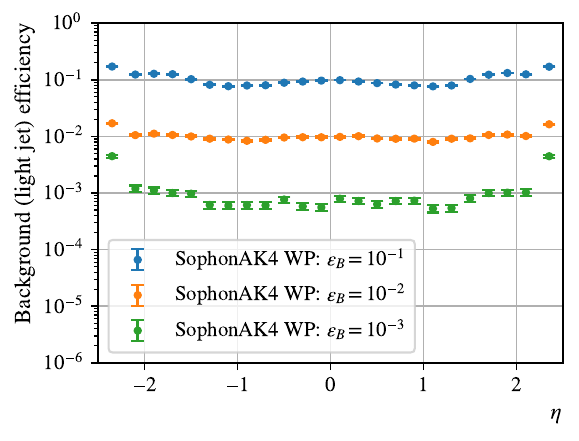}
\caption{The $b$ tagging efficiency (left plots) and light-jet mistagging efficiency (right plots) of \texttt{SophonAK4} as functions of jet \pt (top plots) and $\eta$ (bottom plots), evaluated at three different selection working points. These benchmarks are comparable to the performance of the CMS $b$ and $c$ flavor taggers under the same phase-space selection~\cite{CMS-DP-2024-066}, under the three working points: tight, medium, and loose.
\label{fig:eff_sophonak4}}
\end{figure}

Figure~\ref{fig:eff_sophonak4} further compares the performance of $b$ vs.~light jet tagging across different \pt and $\eta$ regions for three working points. These working points correspond to $b$-tagging discriminant thresholds at background efficiencies ($\epsilon_B$) of $10^{-1}$, $10^{-2}$, and $10^{-3}$.  
By comparing these results with the CMS benchmarks (Ref.~\cite{CMS-DP-2024-066}, Figs. 17, 19, 21, 23, 25, 27, 29, and 31), a similar trend is observed: the tagging performance is lower in the low-\pt and high-$|\eta|$ regions but stabilizes at a plateau beyond the turn-on point.
This indicates that the \SophonAKFour tagger exhibits realistic flavor tagging behavior across different \pt and $\eta$ regions.

\subsection{Tagging region definition of \SophonAKFour}

To facilitate simultaneous $b$ and $c$ tagging in this analysis, we adopt the tagging region definition from a recent ATLAS study. Specifically, the tagging regions are defined using the two-dimensional distribution of $b$-tagging and $c$-tagging discriminants, resulting in five exclusive regions: \texttt{B1} and \texttt{B2} as $b$-tagged jet enriched regions, \texttt{C1} and \texttt{C2} as $c$-tagged jet enriched regions, and \texttt{N} as the non-tagged region.
Each region thus has different selection efficiencies for genuine $b$, $c$, and light jets. The definitions of these regions are based on the tagging region criteria for the widely-adopted \texttt{DL1r} tagger from ATLAS (see Ref.~\cite{ATLAS:VH}, Fig.~1). We adopt the same phase space to enable a direct comparison with the ATLAS results, lowering the jet \pt threshold to 20\GeV.  

\begin{figure}[htbp]
\centering
\includegraphics[width=.60\textwidth]{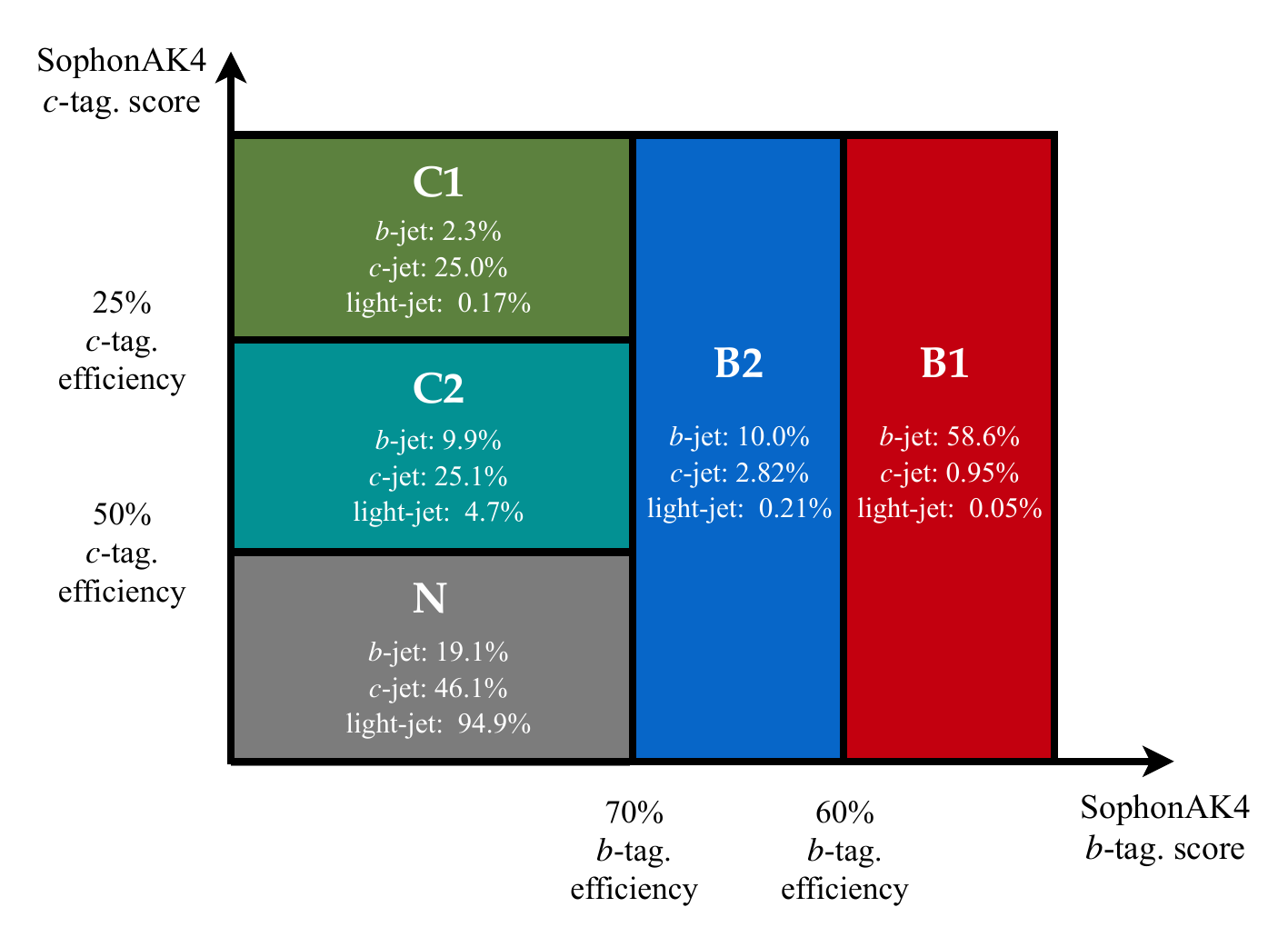}
\vspace{-10pt}
\caption{A schematic of the five flavor tagging regions: \texttt{B1}, \texttt{B2}, \texttt{C1}, \texttt{C2}, and \texttt{N} defined using \SophonAKFour $b$- and $c$-tagging discriminants. The tagging efficiencies annotated for $b$, $c$, and light flavors in these regions are extracted from \ttbar events. The region definitions are based on a recent ATLAS study using the \texttt{DL1r} $b$- and $c$-tagging discriminant (see Fig.~1 of Ref.~\cite{ATLAS:VH}).
\label{fig:region_def_sophonak4}}
\end{figure}

Figure~\ref{fig:region_def_sophonak4} illustrates the five tagging regions and the corresponding selection efficiencies for the three jet flavors. Specifically, in regions \texttt{B1} and \texttt{B2}, the $b$-jet efficiencies are approximately 60\% and 10\%; in regions \texttt{C1} and \texttt{C2}, the $c$-jet efficiencies are about 25\% each. These efficiency values are presented to facilitate a direct comparison with the ATLAS results for the \texttt{DL1r} tagger, as shown in Fig.~1 of Ref.~\cite{ATLAS:VH}.

By comparing the tagging and mistagging efficiencies with the ATLAS \texttt{DL1r} performance~\cite{ATLAS:VH}, we draw to the following findings:
\begin{enumerate}
\item  For $b$ vs.~light jet tagging, the performance is slightly weaker than \texttt{DL1r}: in the \texttt{B1} region, the light-jet efficiency is 0.05\%, consistent with ATLAS; however, in the \texttt{B2} region, the light-jet efficiency is 0.21\%, which is higher than ATLAS's 0.13\%.
\item For $b$ vs. $c$ jet tagging, the performance exceeds that of \texttt{DL1r}: the $c$-jet efficiencies in the \texttt{B1} and \texttt{B2} regions are 0.95\% and 2.8\%, respectively, which are lower than ATLAS's 2.7\% and 5.2\%.
\item For $c$ vs. $b$/light jet tagging, the performance exceeds that of \texttt{DL1r}. In the \texttt{C1} region, the efficiencies for light and $c$ jets are 0.17\% and 2.3\%, respectively, compared to ATLAS's 0.9\% and 4.8\%. A similar trend is observed in the \texttt{C2} region.
\end{enumerate}
These results demonstrate that the \SophonAKFour tagger achieves competitive and, in some cases, superior performance compared to the \texttt{DL1r} tagger, particularly in distinguishing $b$ from $c$ jets and in $c$ vs.~$b$/light jet tagging.  
Considering the comprehensive comparison of \SophonAKFour tagging performance with the benchmarks from ATLAS and CMS taggers, we conclude that \SophonAKFour serves as a reliable and realistic $b/c$ flavor tagger for \DELPHES-based simulations.

\subsection{Scale factors in tagging efficiencies of \SophonAKFour}

To obtain realistic tagging and mistagging efficiency uncertainties based on the five defined tagging regions in this study, we utilize the tagging scale factors derived by ATLAS using 140\invfb of Run-2 data, corresponding to the five ATLAS's tagging regions~\cite{ATLAS:VH}.

For the case of \SophonAKFour, we assume the scale factor central values to be 1, while the uncertainties are assigned according to the ATLAS results for each flavor's tagging and mistagging efficiencies (collectively denoted as $\epsilon_{\rm ftag}$) across the five regions.
The scale factors are jet \pt-dependent. By extracting the relevant results from Ref.~\cite{ATLAS:VH} (Fig.~3), we summarize the $\epsilon_{\rm ftag}$ uncertainties in Table~\ref{tab:sf}. These uncertainties are incorporated into the \Vcb sensitivity estimation conducted in this work.
\begin{table}[!ht]
\centering
\caption{Uncertainties of the scale factors used to correct the \texttt{SophonAK4} tagging efficiencies for $b$, $c$, and light flavors in the five regions: \texttt{B1}, \texttt{B2}, \texttt{C1}, \texttt{C2}, and \texttt{N}, binned by jet \pt. These numbers are extracted from the scale factors and their uncertainties derived from the recent ATLAS scale factor measurements for the \texttt{DL1r} flavor tagger (see Auxiliary material, Fig.~3 of Ref.~\cite{ATLAS:VH}).}
\label{tab:sf}
\vspace{5pt}
\resizebox{\textwidth}{!}{%
\begin{tabular}{cccccccccccccccc}
\hline
    \toprule 
    & \multicolumn{7}{c}{$\epsilon_{\rm ftag}$ uncertainty ($b$ jet)} & \multicolumn{4}{c}{$\epsilon_{\rm ftag}$ uncertainty ($c$ jet)} & \multicolumn{4}{c}{$\epsilon_{\rm ftag}$ uncertainty (light jet)} \\
    \cmidrule(lr){2-8} \cmidrule(lr){9-12} \cmidrule(lr){13-16}
    jet \pt (GeV) & $(20,\,30)$ & $(30,\,40)$ & $(40,\,60)$ & $(60,\,100)$ & $(100,\,175)$ & $(175,\,250)$ & $(250,\,\infty)$ & $(20,\,40)$ & $(40,\,65)$ & $(65,\,140)$ & $(140,\,\infty)$ & $(20,\,50)$ & $(50,\,100)$ & $(100,\,150)$ & $(150,\,\infty)$ \\
    \midrule
    \texttt{B1} & 0.07 & 0.04 & 0.03 & 0.01 & 0.01 & 0.02 & 0.03 & 0.10 & 0.08 & 0.08 & 0.08 & 0.23 & 0.22 & 0.24 & 0.20 \\
    \texttt{B2} & 0.07 & 0.04 & 0.03 & 0.02 & 0.03 & 0.04 & 0.08 & 0.10 & 0.05 & 0.05 & 0.06 & 0.22 & 0.22 & 0.25 & 0.20 \\
    \texttt{C1} & 0.06 & 0.03 & 0.02 & 0.02 & 0.03 & 0.06 & 0.14 & 0.10 & 0.04 & 0.04 & 0.04 & 0.12 & 0.12 & 0.13 & 0.12 \\
    \texttt{C2} & 0.05 & 0.03 & 0.01 & 0.01 & 0.02 & 0.04 & 0.10 & 0.07 & 0.07 & 0.06 & 0.06 & 0.13 & 0.13 & 0.13 & 0.12 \\
    \texttt{N} & 0.20 & 0.12 & 0.08 & 0.05 & 0.07 & 0.08 & 0.13 & 0.07 & 0.04 & 0.04 & 0.04 & 0.0035 & 0.0025 & 0.0025 & 0.0022 \\
    \bottomrule
\end{tabular}%
}
\end{table}

\setcounter{figure}{0}
\setcounter{table}{0}
\section{\boldmath Supplementary details on the \texorpdfstring{\Vcb}{|Vcb|} analysis strategy}\label{app:analysis}

This section provides additional details related to the analysis strategy.
\vspace{10pt}

\paragraph*{Cutflow table.}
For both the resolved and boosted channels, the cutflow is summarized in Table~\ref{tab:cutflow}, following the description provided in the main text. The event yields correspond to those expected for 140\invfb of data. Notice that for the resolved channel, three classifier threshold options (denoted A, B, and C in the table) are considered. Option A is used in the default analysis, while the other option will be detailed below.

\begin{table}[!ht]
\centering
\caption{Cutflow table for boosted and resolved channels. For background processes (\ttbar, \tW and \WW) and signal processes, events are further split into five subcategories based on generator-level matching of the \Wcand large-$R$ jet. Event counts are shown for an integrated luminosity of 140\invfb.}
\label{tab:cutflow}
\vspace{5pt}
\resizebox{\textwidth}{!}{%
\begin{tabular}{p{0.3\linewidth}llllllllllllllllllllll}
    \toprule
    Selection & $W(\ell\nu)$+jets~~ & \multicolumn{6}{c}{\ttbar (semi-lep.)} & \multicolumn{6}{c}{\tW (semi-lep.)} & \multicolumn{3}{c}{\WW (semi-lep.)} & \multicolumn{6}{c}{Signal (\ttbar+\tW+\WW, w/ $W\to bc$ decay)} \\ 
    \cmidrule(lr){3-8} \cmidrule(lr){9-14} \cmidrule(lr){15-17} \cmidrule(lr){18-23}
    ~ & ~ & \textbf{total} & \topbqq & \topbc & \topbq & \Wqq & non & \textbf{total} & \topbqq & \topbc & \topbq & \Wqq & non & \textbf{total} & \Wqq & non & \textbf{total} & \topbqq & \topbc & \topbq & \Wqq & non \\[10pt]
    \midrule
    \multicolumn{23}{c}{\textbf{Boosted channel}} \\ 
    \midrule
    Pass $1\ell$ trigger & 2500000000 & 18000000 & --- & --- & --- & --- & --- & 1600000 & --- & --- & --- & --- & --- & 2300000 & --- & --- & 18000 & --- & --- & --- & --- & --- \\ 
    ~~+~$\geq 1$ exclusive large-$R$ jet & 8300000 & 2900000 & 140000 & 120000 & 520000 & 580000 & 1500000 & 210000 & 4900 & 5700 & 17000 & 62000 & 120000 & 140000 & 38000 & 100000 & 2700 & 120 & 210 & 330 & 530 & 1500 \\ 
    ~~+~$60 < \msd < 110\GeV$ & 1100000 & 1200000 & 17000 & 72000 & 290000 & 480000 & 320000 & 87000 & 640 & 3300 & 9800 & 51000 & 22000 & 46000 & 31000 & 15000 & 1100 & 16 & 120 & 180 & 430 & 310 \\ 
    ~~+~\Sophon \Dbc thres. & 370 & 7600 & 100 & 6400 & 300 & 80 & 720 & 430 & 1.7 & 300 & 9.5 & 10 & 110 & 9.3 & 2.1 & 7.2 & 120 & 0.39 & 11 & 2.3 & 110 & 1.3 \\ 
    ~~+~boosted-chn. classifier thres. & $<$2 & 25 & $<$0.6 & 7.9 & 1.2 & 11 & 4.8 & 0.86 & $<$0.9 & $<$0.9 & $<$0.9 & 0.86 & $<$1.0 & $<$1.0 & $<$1.0 & $<$1.0 & 15 & $<$0.04 & 1.2 & $<$0.04 & 14 & 0.10 \\[10pt]

    \midrule
    \multicolumn{23}{c}{\textbf{Resolved channel}} \\ 
    \midrule
    Pass $1\ell$ trigger & 2500000000 & 18000000 & ~ & ~ & ~ & ~ & ~ & 1600000 & ~ & ~ & ~ & ~ & ~ & 2300000 & ~ & ~ & 18000 & ~ & ~ & ~ & ~ & ~ \\ 
    ~~+~$\geq 4$ exclusive jets \& $N_{\ell}=1$ & 21000000 & 11000000 & ~ & ~ & ~ & ~ & ~ & 580000 & ~ & ~ & ~ & ~ & ~ & 200000 & ~ & ~ & 9600 & ~ & ~ & ~ & ~ & ~ \\ 
    ~~+~$\geq 3$ $b/c$-tagged jets & 210000 & 2300000 & ~ & ~ & ~ & ~ & ~ & 81000 & ~ & ~ & ~ & ~ & ~ & 3400 & ~ & ~ & 5100 & ~ & ~ & ~ & ~ & ~ \\ 
    ~~+~resolved-chn. classifier thres. A & $<$2 & 36 & ~ & ~ & ~ & ~ & ~ & 1.7 & ~ & ~ & ~ & ~ & ~ & $<$2 & ~ & ~ & 26 & ~ & ~ & ~ & ~ \\
    ~~+~resolved-chn. classifier thres. B & $<$2 & 140 & ~ & ~ & ~ & ~ & ~ & 5.2 & ~ & ~ & ~ & ~ & ~ & $<$2 & ~ & ~ & 63 & ~ & ~ & ~ & ~ & ~ \\ 
    ~~+~resolved-chn. classifier thres. C & $<$2 & 270 & ~ & ~ & ~ & ~ & ~ & 10 & ~ & ~ & ~ & ~ & ~ & $<$2 & ~ & ~ & 110 & ~ & ~ & ~ & ~ & ~ \\ 

    \bottomrule
\end{tabular}%
}
\end{table}

\vspace{10pt}

\paragraph*{Event classifier training.}
For the classifier training in both the resolved and boosted channels, the input objects, along with their respective features and vectors, are summarized in Table~\ref{tab:evtdnn_feats}. Both classifiers are trained using deep neural networks based on the Particle Transformer (ParT) architecture. The embedded dimension and MLP layer sizes are reduced by half compared to the standard ParT model.
The classifiers are trained on specifically simulated SM events, with the event categories described in the main text. Care is taken to ensure that the training and validation samples do not overlap with the inference samples used for event yield estimation.
For both channels, the training datasets consist of around 2~M samples.  
The model is trained using a batch size of 512 and an initial learning rate of $10^{-3}$. The dataset is split into 80\% for training and 20\% for validation. The training process spans 20 epochs, with each epoch processing about 0.5~M samples. The optimizer and learning rate scheduler are consistent with those used in the default ParT model training. 

\begin{table}[!ht]
\centering
\caption{Input objects features and vectors for the Particle Transformer event classifier in the boosted- and resolved-channel strategies.}
\label{tab:evtdnn_feats}
\vspace{3pt}
\begin{tabular}{cp{0.7\linewidth}}
    \toprule
     \multirow{4}{*}{\textbf{Objects ($N_\text{tokens}$)~~~}} & \textbf{Resolved channel:} the triggered lepton (1), \ptmiss (1), small-$R$ jets exclusive to the triggered lepton (up to 6) \\[3pt]
   & \textbf{Boosted channel:} the triggered lepton (1), \ptmiss (1), the \Wcand large-$R$ jet (1), small-$R$ jets exclusive to the triggered lepton and \Wcand (up to 5) \\
     \midrule
     \multirow{4}{*}{\textbf{Object features~~~}} & \textbf{Resolved channel:} $\log(\pt)$, $\log(E)$, \texttt{flag}(is triggered lepton), \texttt{flag}(is \ptmiss), \texttt{flag}(is jet), \texttt{flag}(is \texttt{B1}-tagged jet), \texttt{flag}(is \texttt{B2}-tagged jet), \texttt{flag}(is \texttt{C1}-tagged jet), \texttt{flag}(is \texttt{C2}-tagged jet), \texttt{flag}(is \texttt{N}-tagged jet) \\[3pt]
       & \textbf{Boosted channel:} additionally including \texttt{flag}(is large-$R$ jet)  \\
     \midrule
     \multirow{2}{*}{\textbf{Object vectors~~~}} & \textbf{Both channel:} four-vectors $p_i$, used to compute pairwise variables: 
     $m^2_{ij}$,
     $\Delta R_{ij}$,
     $z_{ij} = \frac{\min(p_{\text{T},i}, p_{\text{T},j})}{p_{\text{T},i} + p_{\text{T},j}}$, 
     $k_{\text{T},ij} = \min(p_{\text{T},i},p_{\text{T},j})\,\Delta R_{ij}$
     \\
     \bottomrule
\end{tabular}
\label{tab:tables}
\end{table}

\vspace{10pt}

\paragraph*{Auxiliary uncertainty schemes.}
In estimating the precision of \Vcb, it is important to account for the impact of $b/c$ tagging efficiency uncertainties, which significantly limit the sensitivity of the resolved channel. In this supplementary study, we consider two additional scenarios for these uncertainties. Note that, in the default scenario, we assume that the uncertainties remain constant regardless of the integrated luminosity. This assumption is motivated by the CMS approach, where scale factors are measured separately for each data-taking year, leading to no reduction in statistical uncertainty despite increasing luminosity.  

\begin{figure}[htbp]
\centering
\includegraphics[width=.40\textwidth]{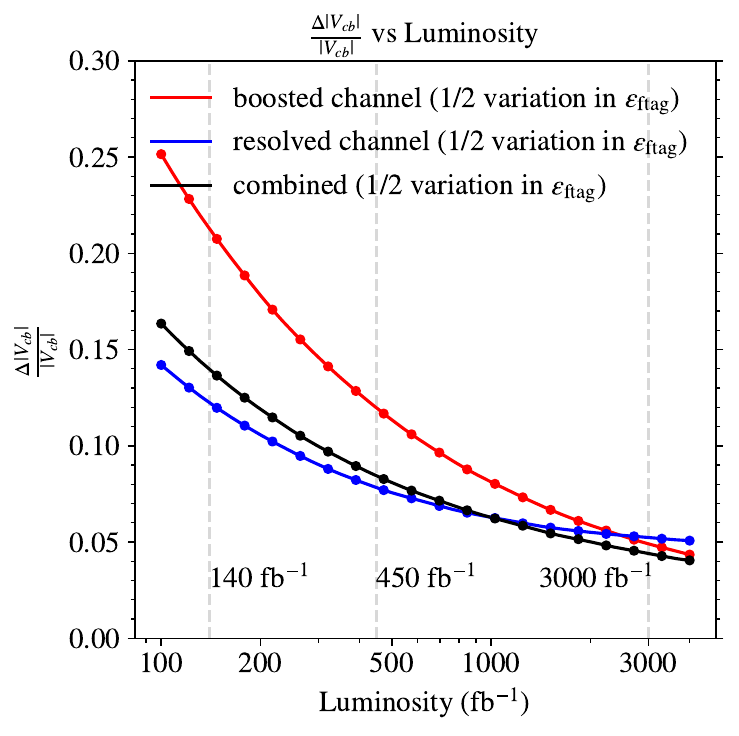}
\includegraphics[width=.40\textwidth]{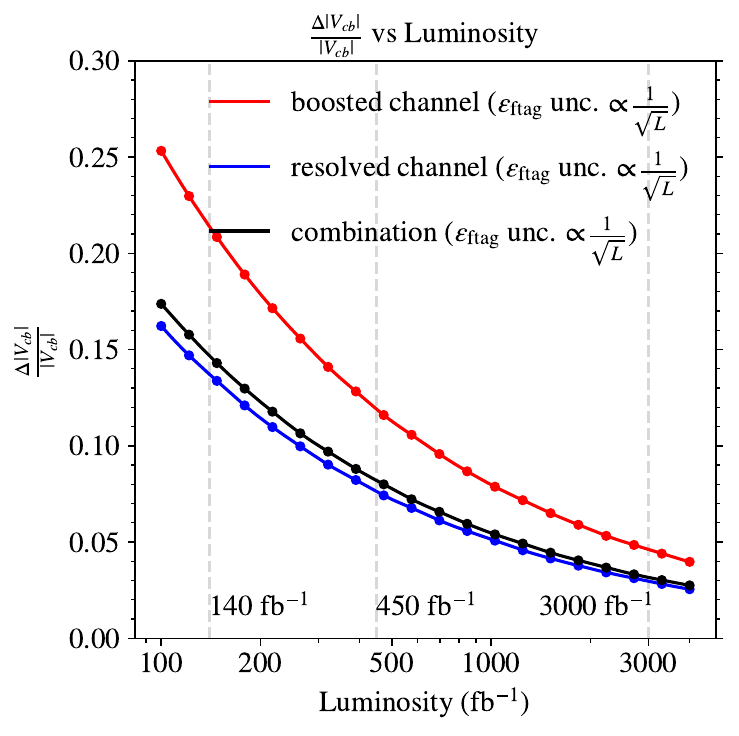}
\caption{The total fractional uncertainty of \Vcb obtained from the boosted and resolved channels, as well as their combination, presented under different luminosity conditions, is shown for two optimistic assumptions regarding the uncertainties in $b/c$ flavor tagging efficiency. \textbf{Left}: Flavor tagging uncertainties are uniformly reduced by a factor of 2, independent of luminosity. \textbf{Right}: Flavor tagging uncertainties are assumed to scale with luminosity as $1/\sqrt{L / 140\invfb}$, representing an overly optimistic scenario at high luminosities.
\label{fig:results_app}}
\end{figure}

In the first scenario, we reduce all $b/c$-(mis)tagging efficiency uncertainties by a factor of 2. This approach allows us to systematically investigate the impact of uncertainty reduction on \Vcb precision. The results are summarized in Fig.~\ref{fig:results_app} (left), showing the precision for the boosted channel, resolved channel, and their combination under this assumption. Notably, the sensitivity of the resolved channel is significantly improved, reaching a level comparable to the boosted channel at 3000\invfb. Comparing the combined results with the conventional one, the improvement in precision reduces from 30\% to 15\%.
It is worth noting that under this assumption, we re-optimized the event classifier thresholds for both the resolved and boosted channels. For the resolved channel, the change in threshold has a substantial impact on event yields, corresponding to ``option B'' in Table~\ref{tab:cutflow}.

In the second scenario, we assume that the default uncertainty values correspond to the 140\invfb dataset and may scale with integrated luminosity $L$ as $1/\sqrt{L}$. Thus, at $L = 3000\invfb$, all uncertainties scale by a factor around 0.2. This scenario reflects an idealized case where a unified tagging scale factor is derived for each experiment using the entire dataset. If the uncertainty is dominated by statistical components, such scaling behavior would be expected.
It is important to emphasize, however, that this scenario is overly optimistic and is considered only as an auxiliary benchmark. In practice, adopting a unified scale factor derived from the entire dataset is not feasible due to time-dependent variations in the collision energy, pileup conditions and the detector performance, etc. As the calibration of $b/c$-(mis)tagging efficiencies is not performed \textit{in situ}, a calibration-by-period strategy is generally more reliable. Even if a unified scale factor is adopted, systematic effects would become dominant at high luminosity, making the estimated 0.2 scaling at 3000\invfb overly optimistic.
Figure~\ref{fig:results_app} (right) summarizes the \Vcb precision under this scaling assumption. In this case, the resolved channel achieves better sensitivity than the boosted channel. The classifier threshold is further relaxed after re-optimization, corresponding to ``option C'' in Table~\ref{tab:cutflow}.

\setcounter{figure}{0}
\setcounter{table}{0}
\section{\boldmath Supplementary details on the \texorpdfstring{$H^{\pm}\to bc$}{H±→bc} search strategy}\label{app:hbc_analysis}

The search for $H^{\pm} \to bc$ from $t\bar{t} \to bH^{\pm} b W^{\mp}$ events follows a similar approach to the search for $W \to bc$ decays in the \ttbar events. This section outlines the technical details of the $H^{\pm} \to bc$ search in the boosted regime. The specific adaptations compared to the strategy for \Vcb extraction are summarized below.
\vspace{10pt}

\paragraph*{Signal process generation.}
The signal process under consideration is $t\bar{t} \to bH^{\pm} b W^{\mp}$ with $H^{\pm} \to bc$, following the ATLAS convention in Ref.~\cite{ATLAS:2023bzb}. The samples are generated using \MGvATNLO~v2.9.18 within the framework of a two-Higgs-doublet model, where the $H^{\pm} \to bc$ decay and leptonic $W$ decay are specified at the matrix element level.
As a benchmark, the mass of $H^{\pm}$ is sampled within the range of 60--160\GeV with 10\GeV steps, with a fixed width of 1\GeV, in line with the ATLAS setup. In addition, signal samples are normalized to the same cross section as the inclusive $t\bar{t}$ sample and assume a reference branching fraction of $\mathscr{B}_{\rm ref} = \mathscr{B}(t\to bH^{\pm}) \times \mathscr{B}(H^{\pm} \to bc) = 1\%$. The generated events are subsequently processed through \PYTHIA~8.3 for hadronization and passed through a fast detector simulation and reconstruction using \DELPHES~3.5, employing the JetClass-II \DELPHES configuration.
\vspace{10pt}

\paragraph*{Event selection and classifier development.}
After applying trigger requirements, the leading boosted jet is identified as the ``$H^{\pm}$ candidate'', replacing the ``$W$ candidate'' in the \Vcb analysis. No mass window selection is imposed on the jet soft-drop mass $m_{\text{SD}}$, as this variable is instead utilized to construct the signal extraction template detailed below.

For training the event-level classifier, the category corresponding to $W \to qq'$ jets (\Wqq-matched category) in the \Vcb analysis is replaced with the $H^{\pm}(qq')$-matched category. The training samples are drawn from dedicated signal events where $H^{\pm}$ is generated with a variable mass, uniformly sampled in the range 50--170\GeV. Boosted jets truth-matched to the $H^{\pm} \to bc$ decay are assigned to the $H^{\pm}(qq')$-matched category, analogous to the definition of \Wqq-matched jets from the simulated \ttbar sample. The ParT classifier is trained using the same configuration as in the \Vcb analysis.

\begin{figure}[htbp]
\centering
\includegraphics[height=.30\textwidth]{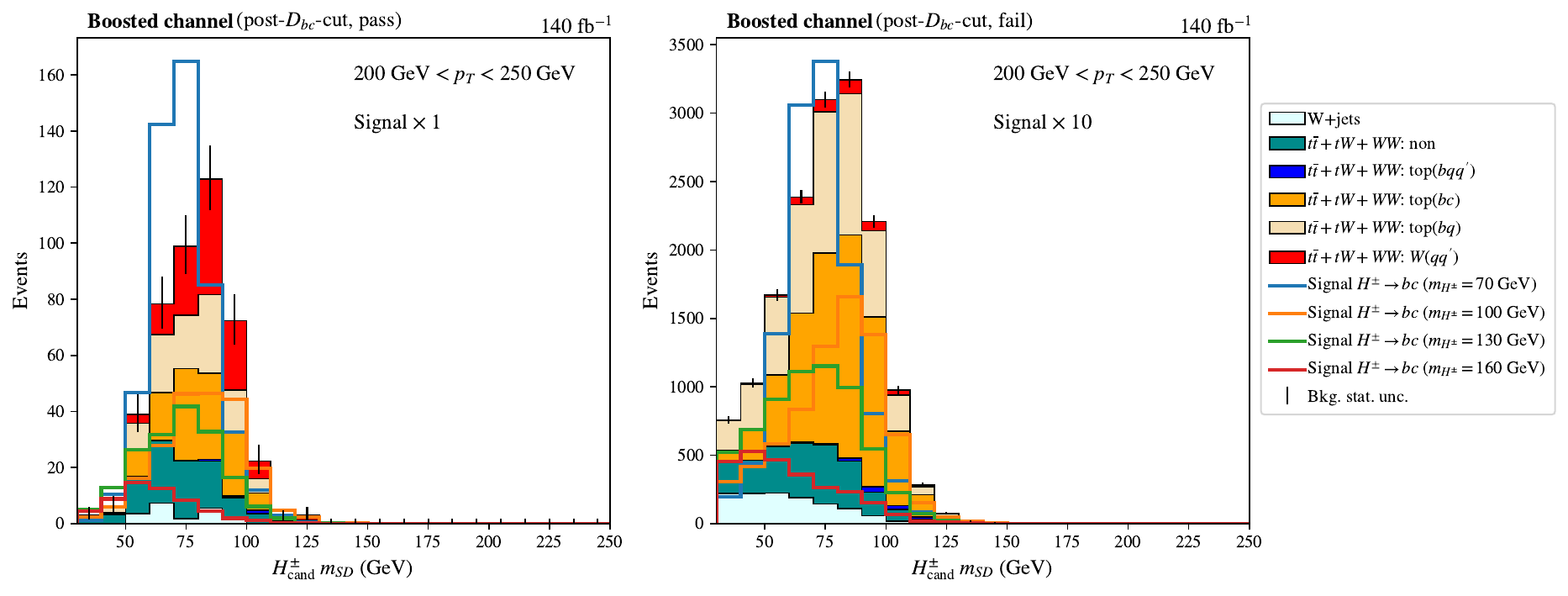}
\includegraphics[height=.30\textwidth]{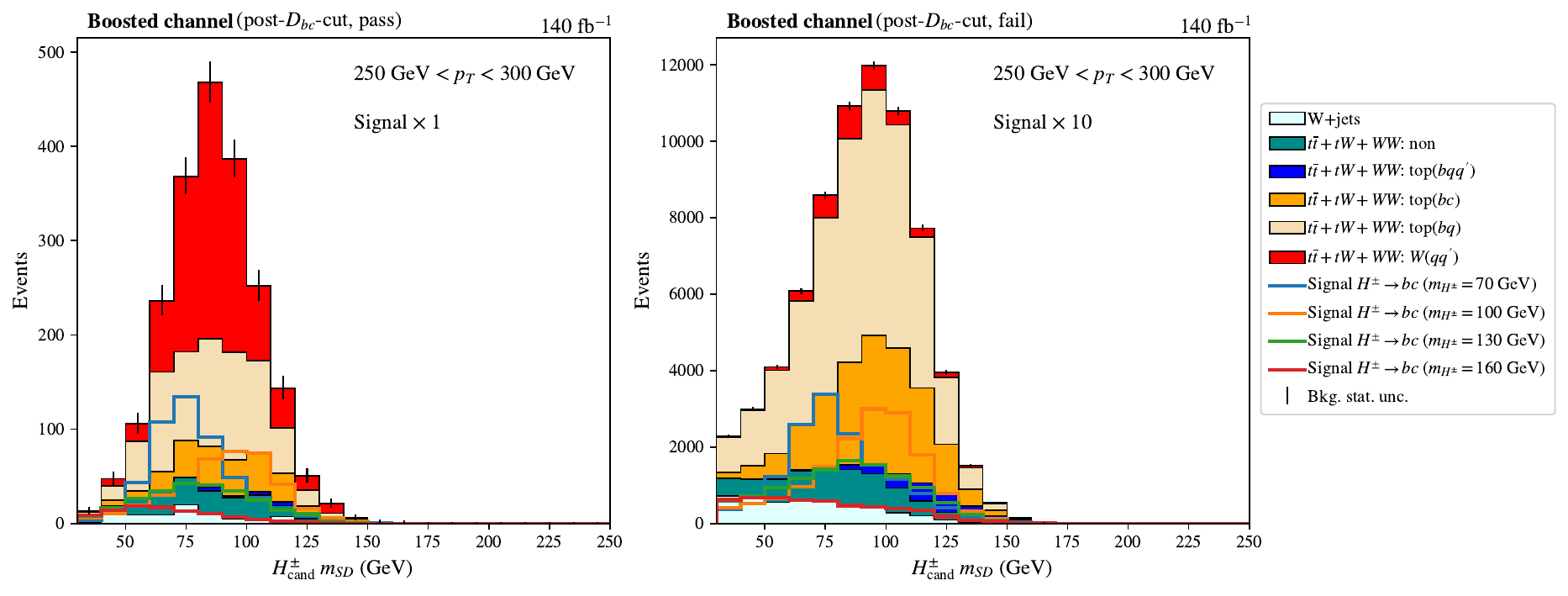} \\
\includegraphics[height=.30\textwidth]{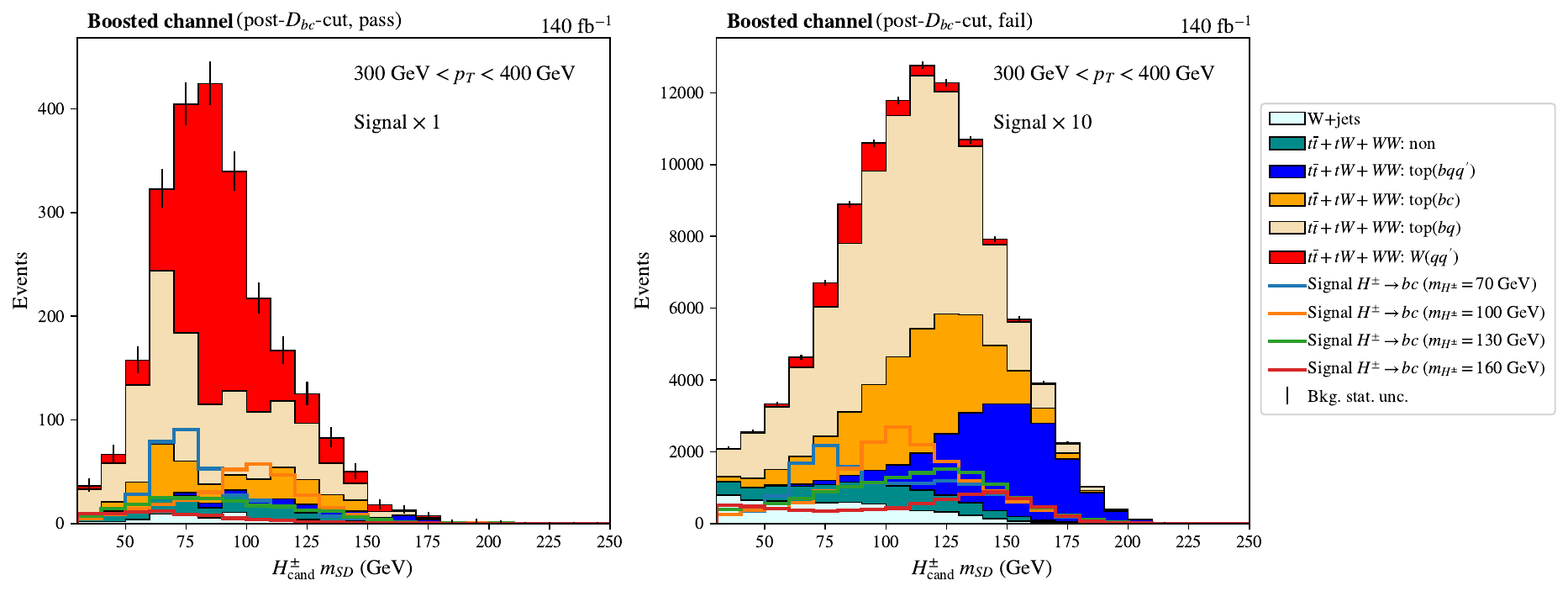}
\includegraphics[height=.30\textwidth]{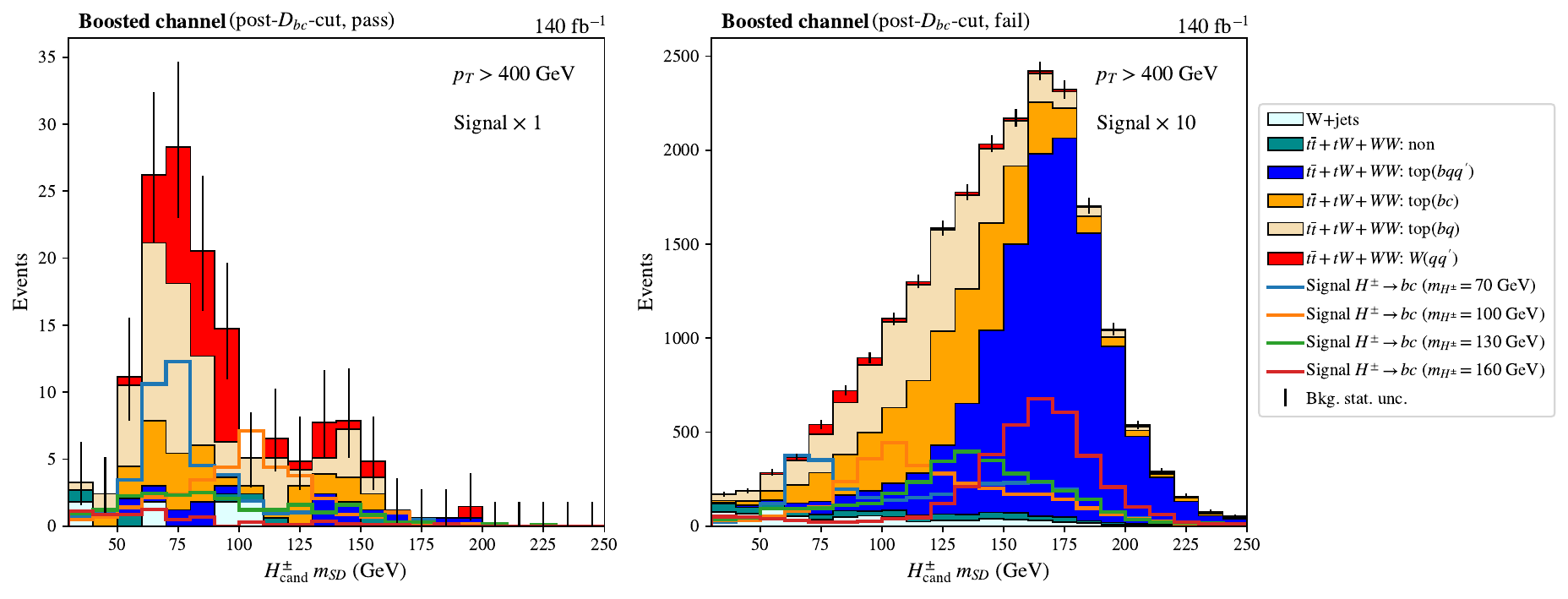} \\
\caption{Distributions of the soft-drop mass \msd of the $H^{\pm}$ candidate jet after applying the optimized \texttt{Sophon} \Dbc selections, shown for the four \pt bins: 200--250, 250--300, 300--400, and $>$400\GeV (from top to bottom). The \texttt{Sophon} \Dbc selection is optimized separately for each \pt region to maximize the sensitivity. For each \pt range, distributions are shown for events passing (left) and failing (right) the optimized event classifier selection.
\label{fig:dist-hbc-msd}}
\end{figure}

We define four analysis regions based on the transverse momentum (\pt) of the $H^{\pm}$ candidate jet, with bins of 200--250, 250--300, 300--400, and $>$400\GeV, to account for different $m_{H^{\pm}}$ signal hypotheses. This binning is motivated by the requirement that higher $m_{H^{\pm}}$ values require higher \pt to ensure that the $bc$ final state is fully reconstructed within the jet cone. For each \pt bin, the \Sophon \Dbc threshold and the event classifier threshold are optimized to maximize the sensitivity for signal extraction from the \msd spectrum. Consequently, the post-\Dbc selection templates for \msd, across the four \pt bins, are shown in Fig.~\ref{fig:dist-hbc-msd}.

\vspace{10pt}

\paragraph*{Fit setup.}
The upper limit on the signal strength, $\mu$, for each mass hypothesis is derived via a binned template fit to the \msd spectrum. The statistical inference is performed using the $\text{CL}_{s}$ method, with the test statistic defined as the profile likelihood ratio. Systematic uncertainties in $b/c$ flavor tagging are incorporated through 15 $\nu$ nuisance parameters, following the same approach as the \Vcb measurement. These parameters account for variations in tagging efficiencies of genuine $b/c$ jets and mistagged light jets across the five \SophonAKFour regions. The uncertainty in the \Sophon \Dbc tagging efficiency is constrained through an \textit{in-situ} calibration, parameterized by a shared scaling factor, $\lambda$, which correlates the efficiency between $bc$-matched background jets and signal jets.

For each \pt bin, where the \Dbc threshold is optimized, events both passing and failing the event classifier selection are included in the analysis, with the signal and background components sharing a correlated $\lambda$ parameter. Since tagging efficiency scale factors vary with different selection thresholds, a distinct $\lambda$ parameter is assigned to each \pt bin. The final signal extraction is performed through a simultaneous fit on the \msd distribution across the four \pt bins, each containing regions passing and failing the event classifier, with a set of nuisance parameters including 15 $\nu$ parameters and four $\lambda$ parameters.

\clearpage

\twocolumngrid
\bibliography{main.bib}

\end{document}